\documentclass[12pt]{article}
\usepackage{amsmath}
\usepackage{amssymb}%

\usepackage{xcolor}
\usepackage{amsmath}
\usepackage{geometry}
\usepackage{mathtools, nccmath}
\usepackage{graphicx}
\usepackage{authblk}

\usepackage{amsmath}
\usepackage{amssymb}
\usepackage{xcolor}
\usepackage[]{lineno}
\usepackage{amsmath}
\usepackage{geometry}
\usepackage{mathtools, nccmath}
\usepackage{graphicx}
\usepackage{hyperref}
\usepackage{xcolor}
\usepackage[]{lineno}
\usepackage{amsmath}
\usepackage{geometry}
\usepackage{mathtools, nccmath}
\usepackage{graphicx}
\usepackage{verbatim}

\title{Stable machine-learning parameterization of subgrid processes for climate modeling at a range of resolutions}

\author{Janni Yuval}
\author{Paul A. O'Gorman}
\affil[1]{Massachusetts Institute of Technology, Cambridge, Massachusetts 02139, USA }

\begin{document}
\date{}
\maketitle



\textbf{Global climate models represent small-scale processes such as clouds and convection using quasi-empirical models known as parameterizations, and these parameterizations are a leading cause of uncertainty in climate projections.
 A promising alternative approach is to use machine learning to build new parameterizations directly from high-resolution model output.
However, parameterizations
learned from three-dimensional model output have not yet been successfully used 
for simulations of climate.  Here
we use a random forest to learn a parameterization of subgrid processes from
output of a three-dimensional high-resolution atmospheric model. 
Integrating this parameterization into the atmospheric model
leads to stable simulations at coarse resolution that replicate the
climate of the high-resolution simulation. The parameterization obeys physical
constraints and captures important statistics such as precipitation extremes.
The ability to learn from a fully three-dimensional simulation
 presents an opportunity for learning parameterizations from
 the wide range of global high-resolution simulations that are now emerging.}
\clearpage

\subsection*{Introduction} 
\hfill

Coupled atmosphere-ocean simulations of climate typically resolve atmospheric processes on horizontal length scales of order 50-100km. 
Smaller-scale processes\textcolor{black}{, such as convection,} are represented by subgrid parameterization
schemes that typically rely on heuristic arguments. Parameterizations are a 
main cause for the large uncertainty in temperature, precipitation and wind projections \cite{webb2013origins,sherwood2014spread,
ogorman2012sensitivity,wilcox2007frequency,ceppi2016clouds,schneider2017climate}.
Although increases in computational resources have now made it possible to run
simulations of the atmosphere that resolve deep convection on global domains for periods of a month or 
more \cite{bretherton2015convective,stevens2019dyamond}, such simulations cannot be run 
for the much longer time scales over which the climate system responds to
radiative forcing \cite{stouffer2004}, and the computational cost to explicitly
resolve important low cloud feedbacks will remain out of reach for the
foreseeable future \cite{schneider2017climate}. 
Therefore, novel and computationally efficient approaches to subgrid parameterization development
are urgently needed and are at the forefront of climate research.

Machine learning (ML) of subgrid parameterizations 
provide one possible route forward
given the availability of high-resolution model output for use 
as training datasets \cite{krasnopolsky2013using,gentine2018could,rasp2018deep,
Ogorman2018using,brenowitz2018prognostic,brenowitz2019spatially,bolton2019applications}.  
\textcolor{black}{While high-resolution simulations still suffer from biases, 
to the extent that they resolve atmospheric convection,
\textcolor{black}{a parameterization learned from these simulations 
has the potential to outperform conventional parameterizations for} 
important statistics such as precipitation extremes.}
Training on both the
control climate and a warm climate is needed to simulate a
warming climate using an ML parameterization \cite{Ogorman2018using,rasp2018deep}, and this is feasible because 
only a relatively short run of a high-resolution model is needed for training data in the warmer climate.  

\textcolor{black}{ML parameterization could also have advantages for grid spacings that are smaller than in current global climate models but not yet convection resolving. At these \textcolor{black}{gray-zone}  grid spacings, assumptions traditionally used in \textcolor{black}{conventional} parameterizations, such as convective quasi-equilibrium
 \cite{arakawa2004cumulus}, may need to be modified or replaced such that the parameterization is \textcolor{black}{scale aware} \cite{arakawa2013unified,ahn2018practical}.  
\textcolor{black}{Without such modifications,  it may be better to turn off some conventional parameterizations of deep convection} for a range of grid spacings that are too close to the convective scale  \cite{pearson2014modelling,vergara2019climate}. 
Since ML parameterizations can be systematically trained at different grid spacings without the need to change physical closure assumptions, an ML approach to parameterization has the potential to perform well across a range of grid spacings and to provide insights into the scale dependence of the parameterization problem.}

Recently a deep artificial neural network (NN) was successfully used to emulate
the embedded two-dimensional cloud-system resolving model in a
superparameterized climate model in an
aquaplanet configuration \cite{gentine2018could,rasp2018deep}, although some choices of NN
architecture \textcolor{black}{could lead to instability and} 
blow ups in the simulations \cite{rasp2019online}.  An NN
parameterization has also been recently learned from the coarse-grained \textcolor{black}{(spatially averaged to a coarser grid)} output
of a fully three-dimensional model, with issues of stability dealt with by 
including multiple time steps in the
training cost function 
and by excluding upper-tropospheric levels from the input
features \cite{brenowitz2018prognostic,brenowitz2019spatially}.  
\textcolor{black}{This} NN parameterization \textcolor{black}{could be used for 
short-term forecasts, but it} suffered from
climate drift on longer times scales and could not be used for studies of
climate. Thus, an ML parameterization has not yet been successfully learned
from a three-dimensional high-resolution atmospheric model for use in studies
of climate.
   
One approach that may help the robustness and stability of an ML
parameterization is to ensure that it respect physical constraints such as
energy conservation \cite{beucler2019achieving}. Using a random forest
(RF) \cite{breiman2001random, hastie2001elements} to \textcolor{black}{learn} a
parameterization has the advantage that the resulting parameterization
automatically respects energy conservation (to the extent energy is linear in
the predicted quantities) and non-negative surface
precipitation \cite{Ogorman2018using}.  \textcolor{black}{An RF is an ensemble of
decision trees, and the predictions of the RF are an average of the predictions
of the decision trees  \cite{breiman2001random,hastie2001elements}.} Physical
constraints are respected by an RF parameterization because the predictions of
the RF are averages over subsets of the training dataset. The property that the
RF predictions cannot go outside the convex hull of the training data may also
help ensure that an RF parameterization is robust when implemented in a global
climate model (GCM).  When an RF was used to emulate a conventional convective
parameterization, it was found to lead to stable and accurate simulations of
important climate statistics in tests with an idealized
GCM \cite{Ogorman2018using}. \textcolor{black}{Thus RFs are promising for use in learning
parameterizations of atmospheric processes, but they have not yet been used to
learn subgrid moist processes from a high-resolution atmospheric model.}

\textcolor{black}{In this study we learn \textcolor{black}{an} RF parameterization from coarse-grained output of 
a high-resolution three-dimensional model of a quasi-global atmosphere, and we
show that the parameterization can be used at coarse
resolution to reproduce the climate of the high-resolution simulation.
By learning different RF parameterizations for a range of coarse-graining factors, we assess the performance of the RF parameterization as the grid spacing in the coarse model is varied,  and this helps addresses the important question of over what range of \textcolor{black}{coarse-graining factors an ML} parameterization of convection can be successful.}

\subsection*{Results} 
\hfill

\subsubsection*{Learning from high-resolution model output}
The model used is the System for Atmospheric Modeling
(SAM) \cite{khairoutdinov2003cloud}, and the domain is an equatorial beta plane
of zonal width $6,912$km and meridional extent $17,280$km 
\textcolor{black}{in an aquaplanet configuration}.
The distribution of sea surface temperature (SST) is specified to be
zonally and hemispherically 
symmetric and reaches a maximum at the equator (the \textcolor{black}{qobs} SST
distribution \cite{neale2000standard}). To
reduce computational expense, we use hypohydrostatic rescaling (with a scaling
factor of 4) which effectively increases the
horizontal length scale of convection and allows us to use a coarser horizontal
grid spacing of 12km than would be normally used in a cloud-system resolving
simulation, while not affecting the large-scale dynamics \cite{kuang2005new,garner2007resolving,boos2016convective,fedorov2019tropical}.
Further details of the model configuration are given in the \textcolor{black}{methods section}.

The high-resolution simulation (hi-res) exhibits organization on a wide range
of length scales from the convective to the planetary scale (Fig. 1a).  The
largest-scale organization consists of two intertropical convergence zones
(ITCZs) and an extratropical storm track in the midlatitudes of each
hemisphere.  The configuration used here in which the SST distribution is fixed
and symmetric about the equator is a challenging test of our RF
parameterization since the resulting circulation is known to be very sensitive
to subgrid parameterizations, and coarse-resolution GCMs in this configuration
give a range of tropical circulations from a strong single ITCZ to a double
ITCZ \cite{mobis2012factors}.  We find there is a double ITCZ at high resolution
for our model configuration, and this is likely dependent on the exact SST
distribution used and the geometry of the domain.  When the model is run with a
horizontal grid spacing of 96km and thus eight times coarser horizontal
resolution (x8), the double ITCZ switches to a much stronger single ITCZ (Fig.
1b) and the distribution of mean precipitation is strongly altered throughout
the tropics (Fig. 2a).  Extreme precipitation, which is important for impacts 
on society and ecosystems, is evaluated here as the 99.9th percentile of 3-hourly precipitation; it is
sensitive at all latitudes to changing from high to coarse resolution (Fig.
2b).  In this study, we do not compare the results of the hi-res simulation
to a coarse-resolution simulation with conventional convective
and boundary-layer
parameterizations both because SAM is not equipped with such
parameterizations
and because the results in the tropics would be highly dependent
on the specific
choice of parameterizations for both mean precipitation \cite{mobis2012factors} and extreme precipitation \cite{wilcox2007frequency}.

\begin{figure}
\centerline{\includegraphics[scale=0.8]{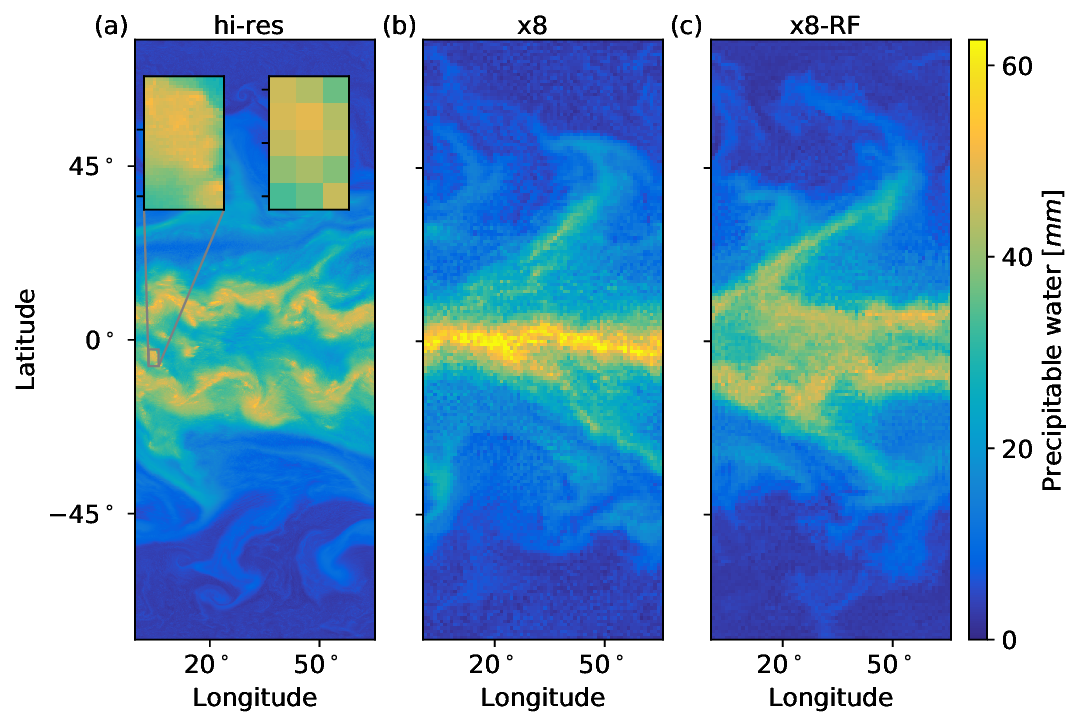}}
\caption{
{\bf Snapshots of column-integrated precipitable water taken from the statistical equilibrium of simulations.} (a) High-resolution simulation (hi-res), (b) coarse-resolution simulation (x8), and (c) coarse-resolution simulation with \textcolor{black}{random forest (RF)} parameterization (x8-RF). Insets in (a) show (left) a zoomed-in region and (right) the same region but coarse-grained by a factor of 8 to the same grid spacing as in (b). The colorbar is saturated in parts of panel b.}
\label{fig:Snapshot_OLR}
\end{figure}

\begin{figure}
\centerline{\includegraphics[scale=0.8]{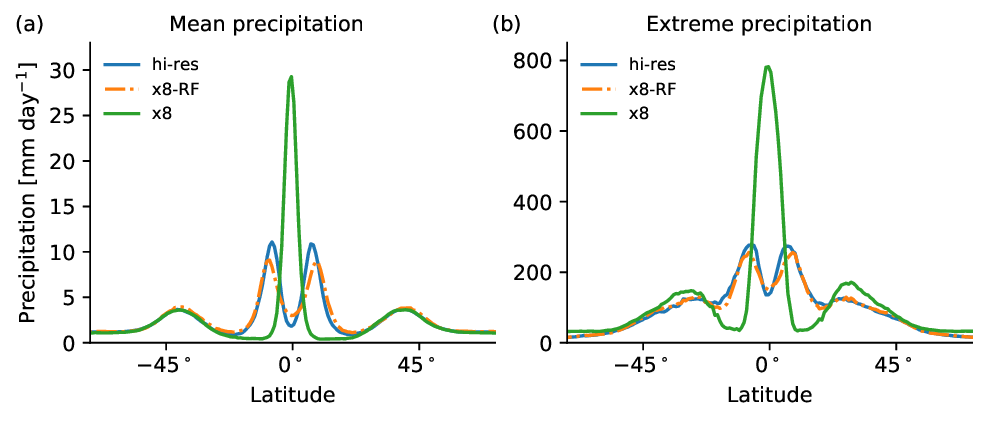}}
\caption{ 
\textbf{Mean and extreme precipitation as a function of latitude.} (a) Zonal- and time-mean precipitation and (b) 99.9th percentile of 3-hourly precipitation, for 
the high-resolution simulation (hi-res; blue),
and the coarse resolution simulation with the \textcolor{black}{random forest (RF)} parameterization (x8-RF; orange dash-dotted) and without the RF parameterization (x8; green).  For hi-res, the precipitation is coarse-grained to the grid-spacing of x8 prior to calculating the 99.9th percentile to give a fair comparison \cite{chen08}.}
\label{fig:precip_online_offline_x8_latin}
\end{figure}

The RF parameterization predicts the effect of unresolved subgrid processes 
\textcolor{black}{that act in the vertical},
including vertical advection, cloud and precipitation microphysics, 
\textcolor{black}{vertical turbulent diffusion, surface fluxes and radiative heating,} on the
resolved thermodynamic and moisture prognostic variables at each grid box and
time step.  
\textcolor{black}{The prognostic variables that
are explicitly affected by the RF-parameterization are the liquid/ice water
moist static energy ($h_{\rm{L}}$), 
\textcolor{black}{
total non-precipitating water mixing ratio ($q_{\rm{T}}$),  
and precipitating water mixing ratio ($q_{\rm{p}}$).}}
\textcolor{black}{Subgrid momentum fluxes are not predicted, but this is
not expected to strongly affect the results since we do not have
 topography that could generate strong gravity wave drag and since tropical 
convection occurs in regions of relatively weak shear in our simulations.}
\textcolor{black}{
We assume that the subgrid contributions depend only on 
the vertical column of the grid point at the current time step,
and we predict all outputs in the vertical column together,
and therefore the parameterization is column based and 
local in time and in the horizontal.
We chose to use two RFs so that we can \textcolor{black}{separately predict}  
processes (turbulent diffusion and surface fluxes) that
depend on horizontal winds and are primarily active at lower levels of the atmosphere.}

\textcolor{black}{The first RF, referred to as RF-tend,  predicts the
vertical profiles at all $48$ model levels of \textcolor{black}{the combined} tendencies due to subgrid vertical advection, 
subgrid \textcolor{black}{cloud} microphysics, subgrid sedimentation and falling of precipitation, and total radiative heating. 
Hence the outputs of RF-tend are $Y_{\rm{RF-tend}} = (h_{\rm{L}}^{\rm{subg-tend}},q_{\rm{T}}^{\rm{subg-tend}},q_{\rm{p}}^{\rm{subg-tend}})$ where subg-tend refers to the subgrid tendency, giving $48 \times 3 =144$ outputs. 
Radiative heating is treated as entirely subgrid, whereas the other processes
have a resolved representation on the coarse model grid and a subgrid component
represented by the RF parameterization.
We do not use the RF-parameterization to predict radiative heating for levels above $11.8$km because it
does not predict the radiative heating well for those levels, possibly because
of insufficient coupling between the stratosphere and troposphere.
 Subgrid tendencies for vertical advection
and microphysics are calculated as the horizontal coarse-graining of the
tendencies at high resolution minus the tendencies calculated from the model
physics and dynamics using the coarse-grained prognostic variables as inputs \textcolor{black}{(see methods)}.
The features (inputs) for RF-tend ($X_{\rm{RF-tend}}$) are chosen 
 to be the vertical profiles (discretized on model levels) of the resolved temperature ($T$), $q_{\rm{T}}$, $q_{\rm{p}}$, and the distance from 
 the equator ($|y|$).  
Hence $X_{\rm{RF-tend}} =(T,q_{\rm{T}},q_{\rm{p}},|y|)$, giving $48 \times 3 + 1 = 145$ features.
Distance from the equator serves as a proxy for the SST, surface albedo and solar insolation, as these are only a function of this distance in the simulations considered here. For a different simulation setup that is not hemispherically symmetric, we would include these physical quantities as separate features instead of distance from the equator.}

\textcolor{black}{The second RF, referred to as RF-diff, predicts the coarse-grained turbulent
diffusivity ($\overline{D}$) for thermodynamic and moisture variables and the
 subgrid correction to the surface fluxes. 
We predict $\overline{D}$ rather than the turbulent diffusive tendencies so as to ensure that the turbulent fluxes remain downgradient. 
For computational efficiency we only predict $\overline{D}$ in the lower troposphere
(the 15 model levels below $5.7$km)
because it decreases in magnitude with height \textcolor{black}{(Supplementary Figure~1d)}.  Hence the outputs 
 of RF-diff are $Y_{\rm{RF-diff}} = (\overline{D}, h_{\rm{L}}^{\rm{surf-flux}},q_{\rm{T}}^{\rm{surf-flux}})$ where surf-flux refers to a \textcolor{black}{subgrid} surface flux, giving $15+1+1=17$ outputs.
 The features of RF-diff are chosen to be the lower tropospheric vertical profiles of $T$, $q_{\rm{T}}$, zonal wind ($u$), meridional wind ($v$),
surface wind speed (wind$_{\rm{surf}}$), and distance from the equator, so that
$X_{\rm{RF-diff}} = (T,q_{\rm{T}},u,v,\rm{wind_{\rm{surf}}},|y|)$, giving
$4\times15+1 + 1 =62$ features.}
\textcolor{black}{Since the meridional velocity
is statistically anti-symmetric with respect to reflection \textcolor{black}{about} the equator,
the meridional wind in the southern hemisphere is multiplied by $-1$ when it is
taken as a feature for RF-diff to help ensure that RF-diff is not learning 
non-physical relationships between inputs and outputs that could artificially improve our results.  
We include the wind variables as features for RF-diff because they improve the prediction of the diffusivity and subgrid surface fluxes. Adding wind features to RF-tend does not improve the accuracy of the predicted tendencies. 
}

The \textcolor{black}{methods section} gives further details about the RFs.
\textcolor{black}{In Supplementary Note~1 we} demonstrate that the \textcolor{black}{RF parameterization} respects the physical constraints of energy conservation \textcolor{black}{(Supplementary Figure~2)} and non-negative surface precipitation \textcolor{black}{(Supplementary Figure~3)}.

\subsubsection*{Simulation with RF parameterization.}
A simulation with the RF parameterization at 96km grid spacing (x8-RF) was run
using an initial condition taken from the statistical equilibrium of the x8
simulation with no RF parameterization. The x8-RF simulation 
transitions to a new statistical equilibrium with a
double ITCZ similar to that in the high-resolution simulation (Fig.~1c)
and it runs stably over long timescales (we have run it for a 1000 days).
At statistical equilibrium, the distribution of mean precipitation is close to
that of the high-resolution simulation (Fig. 2a), and the
distribution of extreme precipitation is remarkably well captured (Fig.
2b).  Other measures such as eddy kinetic energy, mean zonal wind, mean meridional wind
and mean $q_{\rm{T}}$ are also correctly captured by x8-RF \textcolor{black}{(Supplementary Table~1)}.  
Overall, these results show that using the RF subgrid parameterization brings the climate of
the coarse-resolution simulation into good agreement with the climate of the
high-resolution simulation.

The x8-RF simulation requires roughly 30 times less processor time than the
high resolution simulation (for x16-RF the speed up is by roughly a factor of
120).  Further increases in speed could be obtained by increasing the time step
but this is limited \textcolor{black}{in part} by the fall speed of precipitation.  In \textcolor{black}{Supplementary Note~2}, we present an alternative RF parameterization in which
$q_{\rm{p}}$ is no longer treated as a prognostic variable and which could be used to
achieve even faster simulations at coarse resolution \textcolor{black}{in future work}. 
This alternative parameterization \textcolor{black}{has comparable performance  to our
default parameterization \textcolor{black}{(Supplementary Figure~4)} \textcolor{black}{but it requires certain outputs to be set to zero 
(above 11.8km) to avoid a deleterious feedback possibly related to  an 
issue of causality when $q_{\rm{p}}$ is not evolved forward in time \cite{brenowitz2019spatially}, 
and it is} less accurate for extreme precipitation in mid-latitudes.}

\subsubsection*{Performance for different horizontal grid spacings.}
The fact that the RF parameterization is learned from a fully three-dimensional
simulation with a wide range of length scales allows us to explore the
question of whether there is a particular range of grid spacings for
which \textcolor{black}{an ML} parameterization could be most successful.
With increasing grid spacing, coarse-graining involves more averaging over
different cloud elements which should make the subgrid tendencies more
predictable, but the parameterization is then also responsible for more of the
dynamics and physics.  

We train RF parameterizations for a range of coarse-graining factors from x4 to
x32 and use them in simulations with corresponding grid spacings. We first
describe the performance of the RFs on offline tests (i.e., when the RFs are
not implemented in SAM) based on data withheld in training. The offline
performance as measured by the coefficient of determination ($R^2$) improves
substantially as the grid spacing increases (Fig.~3a, compare Fig.~3c and 3e),
consistent with the idea of more predictable subgrid tendencies with more
averaging over larger grid boxes. \textcolor{black}Improved offline performance with increasing
grid spacing is shown to hold for all of the predicted outputs in \textcolor{black}{Supplementary Table~2}.

\begin{figure}
\centerline{\includegraphics[scale=0.8]{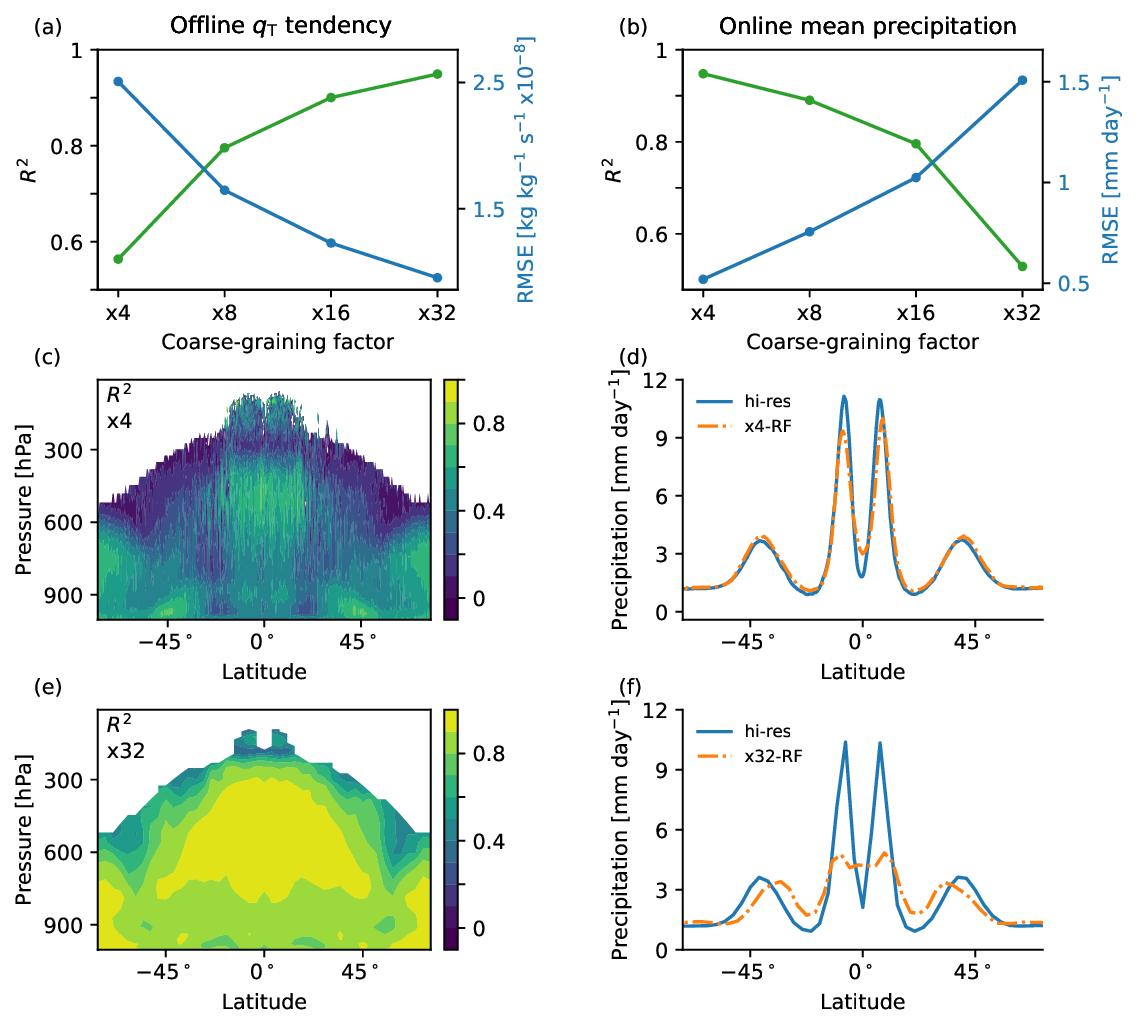}}
\caption{ 
\textbf{Performance of \textcolor{black}{random forest} parameterization versus grid spacing.}
\textcolor{black}{\textcolor{black}{Panels a, c and e} show offline performance as measured on test data for the \textcolor{black}{random forest (RF)} predicted tendency of $q_{\rm{T}}$:} (a) \textcolor{black}{$R^2$ (green) and \textcolor{black}{root mean square error} (blue)
versus grid spacing}, and (c,e) \textcolor{black}{$R^2$} versus pressure and latitude for (c) x4 
and (e) x32.  In (c,e), $\rm{R}^2$ is only shown where the variance is at least $0.1$\% of the mean variance over all latitudes and levels.
\textcolor{black}{\textcolor{black}{Panels b, d and f} show online performance:} (b) $\rm{R}^2$ \textcolor{black}{(green) and \textcolor{black}{root mean square error} (blue)} versus grid spacing \textcolor{black}{for mean precipitation versus latitude}, and (d,f) \textcolor{black}{mean precipitation versus latitude for} hi-res (blue) \textcolor{black}{compared to} (d) x4-RF \textcolor{black}{(orange)} and (f) x32-RF \textcolor{black}{(orange)}.}
\label{fig:Rsq_vs_resolution_qp_Latin}
\end{figure}

\textcolor{black}{However, online performance (i.e., the ability of the coarse-resolution simulations with the RF parameterization to correctly capture the climate of hi-res) 
\textcolor{black}{varies very differently} with grid spacing
as compared to offline performance.
Online performance increases monotonically with decreasing grid spacing (Fig.~3b and compare Fig.~3d and 3f), and the best performance is found for x4-RF, indicating that the RF parameterization can work well with a relatively small gap between the grid spacings of the coarse-resolution \textcolor{black}{model} and \textcolor{black}{the high-resolution model from which it was learned.}}

\textcolor{black}{One might think the decrease in online performance at larger grid spacings is
due to more of the subgrid dynamics and physics becoming subgrid and thus the
absolute errors in the predicted subgrid tendencies becoming larger 
\textcolor{black}{even if $R^2$ increases}, but the
root mean square error (RMSE) in offline tests actually decreases as the grid
spacing increases (Fig.~3a and \textcolor{black}{Supplementary Table~3}).  To understand the
discrepancy between variations in offline and online performance, it is helpful
to think of the variables that the RFs predict as having two components -- a
predictable component and a stochastic component. For smaller grid spacing, the
stochastic component is large (compare 
\textcolor{black}{the same snapshot for different coarse-graining factors in}
Fig~4a and c), and the prediction task becomes
more difficult (compare Fig~4d and f). Therefore, the relatively low offline $R^2$ at smaller grid
spacing does not necessarily imply that the RF does not predict the predictable
component accurately.  To demonstrate this point we make a comparison between
offline performance of x4-RF and x32-RF, but we first coarse grain the subgrid
tendencies calculated and predicted at x4 to the x32 grid, and we refer to the results of this procedure as
x4$\rightarrow$x32 (Fig.~4b \textcolor{black}{and methods}).  The RMSE for x4$\rightarrow$x32 is substantially smaller than for x4 because the
stochastic component averages out with coarse-graining (compare
Fig.~4a and b).
Importantly the RMSE for x4$\rightarrow$x32 is also substantially smaller than
for x32 (compare Fig.~4b and c).  Therefore, x4-RF has smaller offline errors compared to the
x32-RF when these parametrizations are compared in an \textcolor{black}{apples-to-apples}
comparison at the same length scale, and this is consistent with the better
online performance of x4-RF than x32-RF. Similar results are found for other vertical levels and other outputs of the RF \textcolor{black}{(Supplementary Figure~5 and Supplementary Table~4)}. We note that \textcolor{black}{for some outputs}  $R^2$ is still 
\textcolor{black}{higher for x32 than x4$\rightarrow$x32}
\textcolor{black}{(Supplementary Table~4)}, and thus it seems it is more appropriate in
this case to compare the absolute rather than relative errors  of the
parameterization.}

\begin{figure}
\centerline{\includegraphics[scale=0.8]{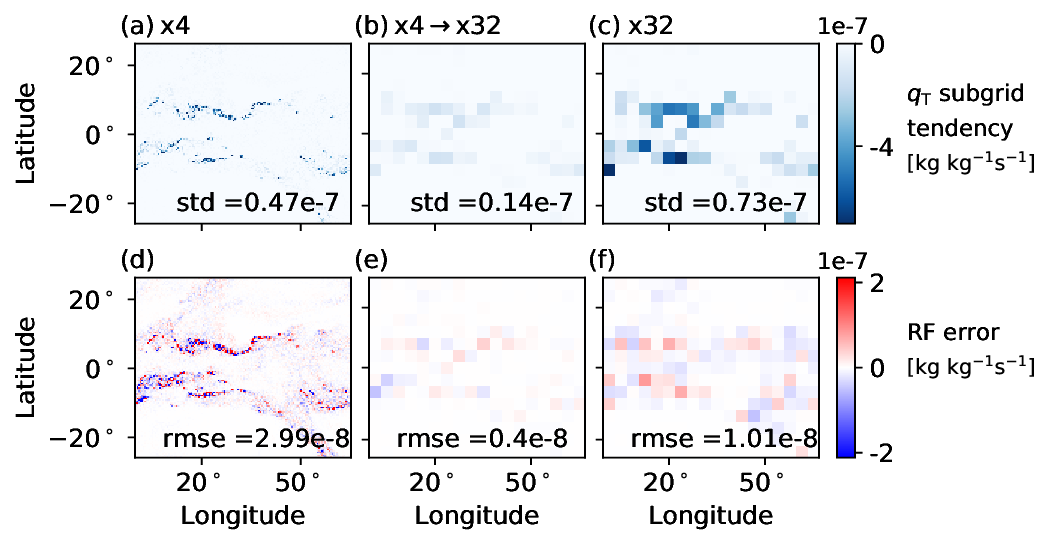}}
\caption{ 
\textbf{\textcolor{black}{Offline comparison of \textcolor{black}{random forest} parameterizations at a common grid spacing.}}
\textcolor{black}{Snapshot of the subgrid tendency of $q_{\rm{T}}$ at $5.7$km \textcolor{black}{in $\rm{kg}~\rm{kg}^{-1} \rm{s}^{-1}$} showing (a-c) the true tendency and (d-f) the error in the prediction from RF-tend. 
Results are shown for (a,d) subgrid tendencies calculated and predicted at x4, (b,e) coarse graining of the subgrid tendencies calculated and predicted at x4 to x32 grid spacing \textcolor{black}{(x4$\rightarrow$x32)}, and (c,f) subgrid tendencies calculated and predicted at x32. Inset text gives for the snapshot shown \textcolor{black}{(a-c)} the standard deviation of the true tendencies \textcolor{black}{and (d-f)} the \textcolor{black}{root mean square error}.
}}
\label{fig:x4_x32_qt_tend_snapshot}
\end{figure}

\textcolor{black}{For extreme precipitation, 
the improvement in online performance with decreasing grid spacing is weaker than for mean precipitation, and there is no improvement in extreme
precipitation performance when decreasing the grid spacing from  x8 to x4
\textcolor{black}{(Supplementary Figure~6)} indicating the possibility of a slight gray zone for this statistic. 
Nonetheless we conclude there is a clear overall improvement in online
performance of the RF parameterization as grid spacing decreases, which suggests that \textcolor{black}{ML} parameterizations could be useful for grid spacings that are quite close to that of the high-resolution model from which they are learned.}

\subsubsection*{Robustness of the RF parameterization}

\textcolor{black}{We performed tests to check the robustness of the RF parameterization, and in particular to confirm that its skill is not based on learning that particular circulations (such as ascent in an ITCZ) and associated clouds occur at particular latitudes.}
\textcolor{black}{As a first test, we re-trained the RF parameterization without using the distance from equator as a feature since much of the subgrid dynamics and physics (e.g., vertical advection, cloud and precipitation microphysics, and longwave cooling) represented by the RF parameterization 
should be largely predictable 
from features other than distance to the equator (which is a proxy for surface albedo, insolation and SST). \textcolor{black}{We find that the offline results are similar regardless of whether distance from the equator is used as a feature in the RFs \textcolor{black}{(Supplementary Tables~2,3)}.}}

\textcolor{black}{As a second test, we trained new versions of the RF
parameterization in which latitudes bands of width $10^\circ$ were excluded
during training in both hemispheres, and thus the RF parameterization must
generalize across latitudes when it is used in simulations. 
 Based on offline tests we find that x8-RF can
generalize remarkably well when tropical latitude bands containing the ITCZs
are excluded (Fig.~5a), and there
is only a slight decrease in performance for excluded latitude bands at high latitudes
(Fig.~5c). 
Excluding latitude bands in mid-latitudes leads to a marked
deterioration in performance 
(Fig.~5b), likely due to the
small overlap between the features 
in the center of the excluded latitude bands
and the training data outside the latitude bands. The lack of 
\textcolor{black}{feature} overlap in the midlatitude case is due 
to strong meridional gradients in
temperature and mixing ratios, and lapse rates and relative 
humidity could be used as alternative features to avoid this overlap problem in future work.
The resulting climates in
coarse-resolution simulations with these RF parameterizations are remarkably
similar \textcolor{black}{(with a slight exception for precipitation in the midlatitude case)} 
to \textcolor{black}{the climate} obtained using x8-RF trained on all the latitudes
(Fig.~5d-f and \textcolor{black}{Supplementary Figure~7}).}

\begin{figure}
\centerline{\includegraphics[scale=0.7]{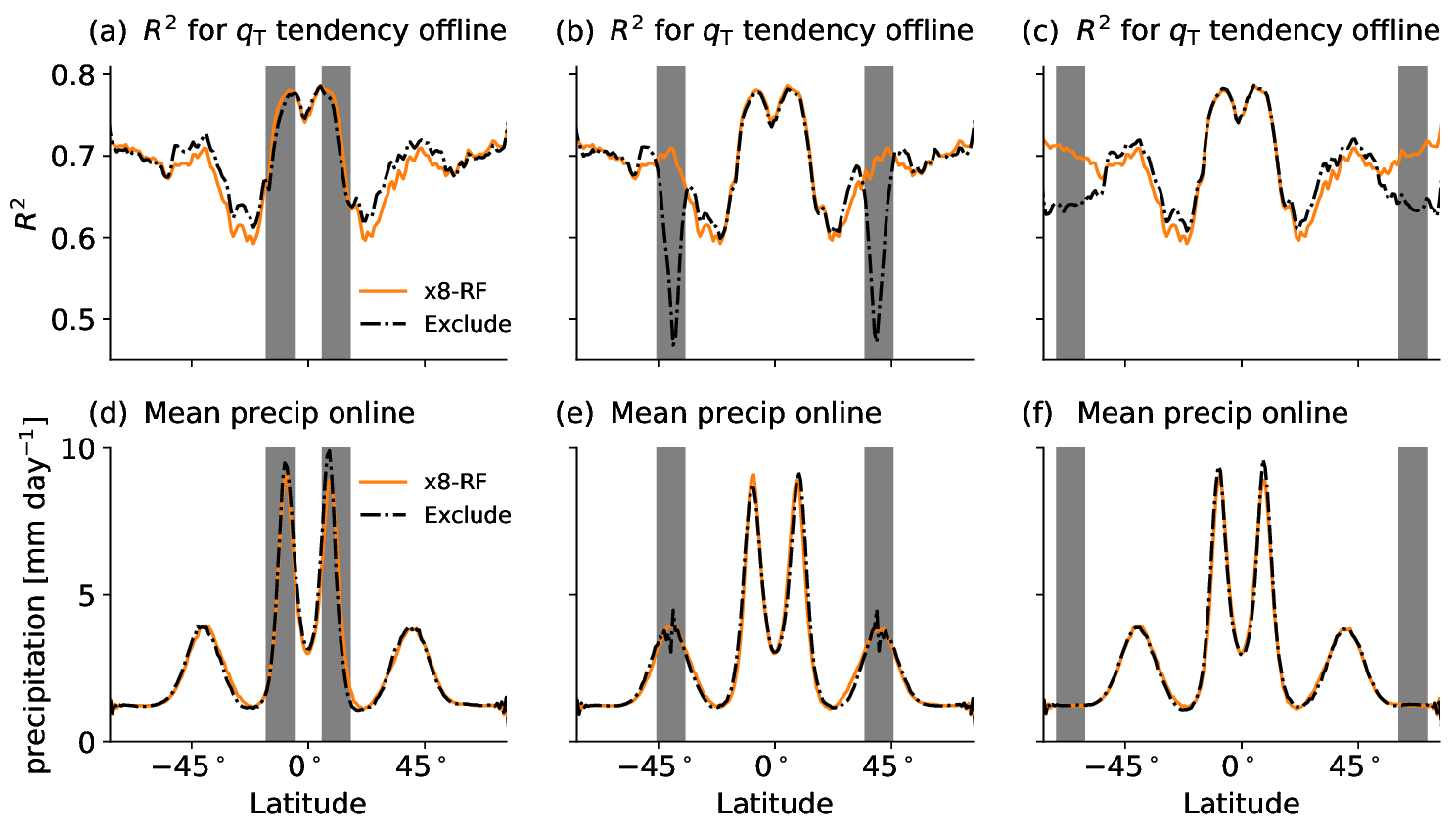}}
\caption{ 
\textbf{\textcolor{black}{Performance \textcolor{black}{of random forest parameterization} when excluding latitude bands during the training process.}} 
\textcolor{black}{(a-c) 
Offline performance at x8 as measured by $\rm{R}^2$ of the tendency of $q_{\rm{T}}$ (including all longitudes and levels) versus latitude, and 
(d-f) online performance as shown by the zonal- and time-mean precipitation for the x8-RF simulation. Results are shown in black dash-dotted for cases in which the training process excludes in both hemispheres the latitude bands
(a,d) $5.1^{\circ}-15.5^{\circ}$,
(b,e)  $34.5^{\circ}-44.9^{\circ}$,
and (c,f) $60.5^{\circ}-70.8^{\circ}$.
Grey bars indicate latitude bands that where excluded during the training.
For comparison the results for x8-RF without any latitudes excluded in training are plotted in orange.}}
\label{fig:Exclude_12_24_40_52_70_82_offline_online}
\end{figure}

\textcolor{black}{ The results of these tests suggest that
the success of the RF parameterization is not based on learning that particular circulation features 
occur at particular latitudes (for example, the RF parameterization is successful even when its training excludes the ITCZ regions), but rather it is learning robust physical relationships between features and outputs.}

\subsection*{Discussion}
\hfill

The results presented here provide a step forward by demonstrating the
viability of stable and accurate parameterizations of subgrid physics and dynamics learned from a
high-resolution three dimensional simulation of the atmosphere. 
\textcolor{black}{The results also give insights into how well \textcolor{black}{an ML}
parameterization can perform as a function of grid spacing.  
Online performance improves with decreasing grid spacing of the coarse-resolution model.  This is in contrast to the experience that \textcolor{black}{some} conventional parameterizations are best turned off
for a range of length scales that are too close to the convective
scale \cite{pearson2014modelling,vergara2019climate}, and the difference may arise because 
 conventional parameterizations 
rely on physical assumptions that are not uniformly valid
across length scales, although this can be mitigated by trying to make such
parameterizations scale aware  \cite{arakawa2013unified,grell2014scale,ahn2018practical}. 
\textcolor{black}{Care is needed in comparing offline performance across grid spacings, and
we find that it is useful to compare offline error statistics at a consistent reference length
scale.}
Further work using a model without hypohydrostatic
scaling would be helpful to further investigate the behavior of \textcolor{black}{ML}
parameterizations at different grid spacings.}

\textcolor{black}{The approach to ML parameterization for the atmosphere in this study is different in important aspects to previous studies.
First, the predicted tendencies are 
calculated accurately for the instantaneous atmospheric state rather than approximating them
based on differences over 3-hour periods  \cite{brenowitz2018prognostic,brenowitz2019spatially}.  Second, the subgrid corrections are calculated independently for each physical process rather 
than for all processes together
as in previous studies \cite{brenowitz2018prognostic,rasp2018deep,brenowitz2019spatially} which allows 
for an ML parameterization structure that is motivated by physics
and the calculation of the precipitation rate
from the predicted tendencies.
Third, we use an RF to learn from a high-resolution model 
whereas NNs have been used in previous studies 
that learned from a high-resolution model
 \cite{krasnopolsky2013using,brenowitz2018prognostic,rasp2018deep,brenowitz2019spatially}
and RFs were used only to emulate conventional parameterizations  \cite{belochitski2011tree,Ogorman2018using}.
Parameterizations based on an RF have advantages in that their predictions automatically satisfy physical properties in the training data
(without being imposed explicitly \cite{beucler2019achieving})
and they make conservative predictions for samples outside of the training data which may help with \textcolor{black}{the robustness of their online performance. On the other hand, NNs require less memory and may have better offline performance}.  To further compare RF and NN parameterizations, future work should evaluate their online and offline performance using the same training data and atmospheric model.} 

Future research on ML parameterization for the atmosphere must
address technical challenges such as how best to train over land regions with topography and \textcolor{black}{how best to deal with the need for a separate parameterization of radiative heating in the stratosphere. However, future research
should also continue to seek insights into 
the nature of the parameterization problem, such as how performance varies
across length scales or whether parameterizations should be nonlocal in time and space, which may also inform the further development of conventional parameterizations.}

\subsection*{Methods}
\subsubsection*{Model}
The model used in this study is the System for Atmospheric Modeling (SAM), version 6.3 \cite{khairoutdinov2003cloud}, which is a relatively efficient model that integrates the anelastic equations of motion in Cartesian coordinates.  
The bulk microphysics scheme is single moment with precipitating water consisting of rain, snow and graupel, and non-precipitating water consisting of water vapor, cloud water and cloud ice.
Cloud ice experience sedimentation, and we include the surface sedimentation flux (which is small) in all reported surface precipitation statistics.
The subgrid-scale turbulent closure is a Smagorinsky-type scheme. 
The radiation scheme is based on parameterizations from the National Center for Atmospheric Research (NCAR) Community Climate Model (CCM) version 3.5 \cite{kiehl1998national}.

The equations for the prognostic  thermodynamic and moisture variables in SAM are important for our study and may be written as \cite{khairoutdinov2003cloud}
\begin{fleqn}
 \begin{equation}
 \frac{\partial h_{\rm{L}}}{\partial t} = -\frac{1}{\rho_{\rm{0}}} \frac{\partial}{\partial x_i}(\rho_{\rm{0}} u_i h_{\rm{L}} + F_{h_{\rm{L}} i}) -\frac{1}{\rho_{\rm{0}}}  \frac{\partial}{\partial z}(L_{\rm{p}} P_{\rm{tot}}+ L_{\rm{n}} S) + \left(\frac{\partial h_{\rm{L}}}{\partial t}\right)_{\rm{rad}},
 \label{eq:hL equation} 
 \end{equation} 
\begin{equation}
\frac{\partial q_{\rm{T}}}{\partial t} = -\frac{1}{\rho_{\rm{0}}} \frac{\partial}{\partial x_i}(\rho_{\rm{0}} u_i q_{\rm{T}} + F_{q_{\rm{T}} i}) +\frac{1}{\rho_{\rm{0}}}  \frac{\partial}{\partial z}(S) -\left(\frac{\partial q_{\rm{p}}}{\partial t}\right)_{\rm{mic}},
 \label{eq:qT equation} 
\end{equation}
\begin{equation}
\frac{\partial q_{\rm{p}}}{\partial t} = -\frac{1}{\rho_{\rm{0}}} \frac{\partial}{\partial x_i}(\rho_{\rm{0}} u_i q_{\rm{p}} + F_{q_{\rm{p}} i}) + \frac{1}{\rho_{\rm{0}}}  \frac{\partial}{\partial z}( P_{\rm{tot}})+\left(\frac{\partial q_{\rm{p}}}{\partial t}\right)_{\rm{mic}},
 \label{eq:qp equation}
\end{equation}
\end{fleqn}
where $h_{\rm{L}}= c_{\rm{p}}T + gz - L_{\rm{c}}(q_c + q_{\rm{r}}) - L_{\rm{s}}(q_{\rm{i}} + q_{\rm{s}} + q_{\rm{g}})$ is the liquid/ice water static energy; $\rho_{\rm{0}}(z)$ \textcolor{black}{is the reference density profile}; $q_{\rm{T}}$ is the non-precipitating water mixing ratio which is the sum of the mixing ratios of water vapor  ($q_v$),  cloud water ($q_c$) and cloud ice ($q_{\rm{i}}$); $q_{\rm{p}}$ is the total precipitating water mixing ratio which is the sum of the mixing ratios of rain ($q_{\rm{r}}$),  snow ($q_{\rm{s}}$) and graupel ($q_{\rm{g}}$); $ F_{Ai}$ is the diffusive flux of variable $A$; $u_i = (u,v,w)$ is the three-dimensional wind;
$P_{\rm{tot}}$ is the total precipitation mass flux (defined positive downwards); $S$ is the total sedimentation mass flux (defined positive downwards); the subscript \textcolor{black}{rad} denotes the tendency due to
radiative heating; the subscript \textcolor{black}{mic} represents the microphysical tendency due to autoconversion, aggregation, collection, and evaporation and sublimation of precipitation;
$L_{\rm{c}}$, $L_{\rm{f}}$ and $L_{\rm{s}}$ are the latent heat of condensation, fusion and sublimation, respectively; $L_{\rm{p}} = L_{\rm{c}} + L_{\rm{f}}(1-\omega_{\rm{p}})$ is the effective latent heat associated with precipitation, and $\omega_{\rm{p}}$ is the partition function for precipitation \textcolor{black}{ which determines its partitioning between liquid and ice phases}; $L_{\rm{n}} = L_{\rm{c}} + L_{\rm{f}}(1-\omega_{\rm{n}})$ is the effective latent heat associated with non-precipitating condensate, and $\omega_{\rm{n}}$ is the partition function for non-precipitating condensate \textcolor{black}{which determines its partitioning between liquid and ice phases. We note that we do not introduce any prescribed large-scale tendencies in our simulations.}

\subsubsection*{Simulations}
All simulations are run on the same quasi-global domain with an equivalent latitude range from $-78.5^\circ$ to $78.5^\circ$ and  longitudinal extent of $62.2^\circ$ at the equator.
There are $48$ vertical levels with spacing that increases from $85$m at the surface to $1650$m in the stratosphere\textcolor{black}{, and the top level is at $28695$m}. The default time step is $24$ seconds, \textcolor{black}{ and this is adaptively reduced as necessary to prevent violations of the CFL condition.}
The insolation is set at perpetual equinox without a diurnal cycle. The simulations are run with a zonally symmetric \textcolor{black}{qobs} \cite{neale2000standard} sea surface temperature (SST) distribution which varies between $300.15\rm{K}$ at the equator and $273.15\rm{K}$ at the poleward boundaries. 
Surface albedo is a function of latitude, and there is no sea ice in the model. 
\textcolor{black}{Simulations with a diurnal cycle and different SST distributions should be investigated in future work.}

Hypohydrostatic rescaling of the vertical momentum equation with a rescaling factor of 4 increases the horizontal length scale of convection while leaving the large-scale dynamics unaffected and still retaining a very large range of length scales in the hi-res simulation \cite{kuang2005new,garner2007resolving,ma2014effects,boos2016convective,fedorov2019tropical}. 
A similar configuration of SAM with hypohydrostatic rescaling (though not at equinox) was recently used to investigate tropical cyclogenesis in warm climates \cite{fedorov2019tropical}.
Furthermore, SAM was also used in previous studies that developed ML parameterizations \cite{rasp2018deep,brenowitz2018prognostic,brenowitz2019spatially}.

The hi-res simulation has $12$km grid spacing (recalling that hypohydrostatic rescaling is used) and was spun up for $100$ days. It was then run for $500$ days with three-dimensional snapshots of the prognostic variables, radiative heating and turbulent diffusivity saved every three hours. Results for the hi-res simulation are averaged over $500$ days.  Coarse-resolution simulations were run for $600$ days, with the first $100$ days of each simulation treated as spinup, and results averaged over the last $500$ days. 
Simulations with the RF parameterization start with initial conditions taken from simulations without the RF parameterization (at the same resolution). 
\textcolor{black}{The transition in the simulations with the RF parameterization from a single ITCZ in the initial condition to a double ITCZ sometimes occurs in two distinct steps, but a spinup period of $100$ days was found to be sufficient for this transition to occur.}

The version of SAM that was used for the hi-res simulation had some minor discretization errors, the most important of which were in the Coriolis parameter in the meridional momentum equation \textcolor{black}{and in the momentum surface fluxes}. Effectively the Coriolis parameter is shifted by a distance of half a gridbox ($6$km) to the south, \textcolor{black}{and the surface winds used for calculating the surface momentum fluxes are also shifted by a distance of half a gridbox (but each wind component in different direction). 
We corrected these errors when running the coarse-resolution simulations since \textcolor{black}{discretization errors} become larger in magnitude with coarser grid spacing.}
To avoid wasteful rerunning of the expensive hi-res simulation, in all coarse-resolution simulations we also shifted the Coriolis parameter in the meridional momentum equation by $6\rm{km}$ (half of the hi-res gridbox size) \textcolor{black}{and shifted the surface winds by $6\rm{km}$ when calculating the surface momentum fluxes} such that the coarse-resolution simulations are completely consistent with the hi-res simulation. 

\subsubsection*{Coarse graining and calculation of subgrid terms \label{sub:Coarse graining}}
For each 3-hourly snapshot from the hi-res simulation, we coarse grain the prognostic variables $(u,v,w,h_{\rm{L}},q_{\rm{T}},q_{\rm{p}})$, the tendencies of $h_{\rm{L}},q_{\rm{T}},$ and $q_{\rm{p}}$ (eqs.~\ref{eq:hL equation}-\ref{eq:qp equation}), the surface fluxes and the turbulent diffusivity. 
Coarse-graining is performed by horizontal averaging onto a coarser grid as follows:
\begin{equation}
\overline{A}(i,j,k) = \frac{1}{N^2}\sum_{l=N(i-1)+1}^{l=Ni} \, \sum_{m=N(j-1)+1}^{m=Nj}A(l,m,k),
\end{equation}
where $A$ is the high-resolution variable, $\overline{A}$ is the coarse-grained variable, $N$ is the coarse graining factor, $k$ is the index of the vertical level, and  $i,j$ ($l,m$) are the discrete indices of the longitudinal and latitudinal coordinates  at coarse resolution (high resolution). 

Different coarse-graining factors were used to study how well the ML-parameterization performs at different resolutions. The horizontal grid spacings that were used were $48\rm{km}$ (x$4$), $96\rm{km}$ (x$8$), $192\rm{km}$ (x$16$), and $384\rm{km}$  (x$32$).  The hi-res simulation has a grid size of $576\rm{x}1440$, and coarse graining it by factors of $4$, $8$ and $16$ results in grid sizes of $144\rm{x}360$,  $72\rm{x}180$ and $36\rm{x}90$, respectively. These grids can be simulated in SAM. Unfortunately, coarse-graining the hi-res simulation by a factor of $32$ results in a grid  ($18\rm{x}45$) which cannot run \textcolor{black}{in} SAM. Instead, the number of grid points in the latitudinal direction in these simulations was increased to $48$ points ($18\rm{x}48$ grid size), leading to a slightly larger domain, and the presented results were interpolated to the coarse-grained high-resolution grid (with $45$ points in the latitudinal direction).

We define the resolved tendency as the tendency calculated using the dynamics and physics of model with the coarse-grained prognostic variables as inputs.
The tendencies due to unresolved (subgrid) physical processes were calculated as the difference between the coarse-grained tendency and the resolved tendency. The subgrid tendency for a given process is then written as
\begin{linenomath*}
\begin{equation}
\left(\frac{\partial \overline{B}}{\partial t}\right)^{\rm{subgrid}} =  
\frac{\partial \overline{B}}{\partial t}(h_{\rm{L}},q_{\rm{T}},q_{\rm{p}},u,v,w) - \frac{\partial B}{\partial t}(\overline{h}_{\rm{L}},\overline{q}_{\rm{T}},\overline{q}_{\rm{p}},\overline{u},\overline{v},\overline{w})
\label{eq:sgs process calc}
\end{equation}
\end{linenomath*}
where $B$ is a certain variable, $\frac{\partial \overline{B}}{\partial t}(h_{\rm{L}},q_{\rm{T}},q_{\rm{p}},u,v,w)$ is the coarse-grained high-resolution tendency of that variable due to the process, $\frac{\partial B}{\partial t}(\overline{h}_L,\overline{q}_T,\overline{q}_p,\overline{u},\overline{v},\overline{w})$ is the resolved tendency due to the process, and  $\left(\frac{\partial \overline{B}}{\partial t}\right)^{\rm{subgrid}}$ is the subgrid tendency due to the process. 
For example, the subgrid tendency of $h_{\rm{L}}$ due to vertical advection is
\textcolor{black}{
\begin{linenomath*}
\begin{equation}
\left(\frac{\partial \overline{h}_{\rm{L}}}{\partial t}\right)_{\rm{vert.\,adv.}}^{\rm{subgrid}} =-\left(\frac{\partial \overline{w h_{\rm{L}}}}{\partial z}- \frac{\partial \overline{w} \overline{h}_{\rm{L}}}{\partial z}\right).
\label{eq:sgs adv calc}
\end{equation}
\end{linenomath*}
}
Subgrid and resolved contributions are defined in a similar way for the surface fluxes of $h_{\rm{L}}$ and $q_{\rm{T}}$.

\textcolor{black}{The procedure of coarse graining and calculating the subgrid tendencies and subgrid surface fluxes was done offline in postprocessing. For each high resolution snapshot, the coarse-grained fields, the instantaneous tendencies associated with different physical processes, and the surface fluxes were calculated. The coarse-grained fields were then used to calculate the instantaneous resolved tendencies of the different physical processes and the resolved surface fluxes. Finally the subgrid contributions were calculated. This procedure is more accurate compared to previous studies that calculated the tendencies using the difference between the prognostic variables over 3-hour time steps \cite{brenowitz2018prognostic,brenowitz2019spatially} . Furthermore, this procedure allows us to calculate a different sub-grid tendency for each physical \textcolor{black}{process}, which is necessary for the RF-parameterization structure \textcolor{black}{that we use.}} 

\subsubsection*{Choice of outputs for the RF parameterization \label{sub:outputs}}
The RF parameterization predicts the combined tendencies for the following processes: subgrid vertical advection of $h_{\rm{L}}, q_{\rm{T}}$, and $q_{\rm{p}}$, subgrid cloud and precipitation microphysical tendencies included in $\left(\frac{\partial q_{\rm{p}}}{\partial t}\right)_{\rm{mic}}$, subgrid falling of precipitation and subgrid sedimentation of cloud ice, and the total radiative heating tendency (see below). The RF parameterization also predicts the \textcolor{black}{coarse-grained turbulent diffusivity and the} subgrid corrections to the surface fluxes of $h_{\rm{L}}$ and $q_{\rm{T}}$.

For radiation, the RF parameterization predicts the total radiative heating and not the subgrid part. The choice to predict the radiative heating tendency rather than predicting its subgrid correction was  mainly motivated by the complexity of calculating subgrid radiative heating tendencies in post-processing. 
Radiative heating is not predicted above $11.8\rm{km}$ since the RF has poor performance above this level in offline tests.  Instead the SAM prediction for radiative heating is used at those levels.  We checked that the results were not sensitive to the exact choice of cutoff level.  Including the RF prediction for radiative heating at all stratospheric levels leads to a temperature drift in the stratosphere when  RF-tend is implemented in SAM \textcolor{black}{(a problem with temepratures in the stratosphere was also found in a previous study \cite{rasp2018deep})}, though tropospheric fields are still similar to the presented results. 
\textcolor{black}{It is possible that due to weak troposphere-stratosphere coupling it is difficult to accurately predict the radiative heating tendency simultaneously in both the troposphere and the stratosphere. In future work, it 
might be beneficial to train different parameterizations for the stratosphere and troposphere.}

\textcolor{black}{The turbulent vertical diffusive flux for
a thermodynamic or moisture variable $A$ is $F_{Az}=-D\frac{\partial A}{\partial z}$, where $D$ is the
turbulent diffusivity for thermodynamic and moisture variables.}
\textcolor{black}{We} predict the coarse-grained turbulent
diffusivity \textcolor{black}{($\overline{D}$)} \textcolor{black}{and apply it only to vertical diffusion of the
thermodynamic and moisture variables (i.e., $h_{\rm{L}}, q_{\rm{T}}, q_{\rm{p}}$).
\textcolor{black}{This is consistent with our general approach in which 
the RF parameterization only represents processes that act in the vertical and only their effects on the
thermodynamic and moisture variables}.  
 The approach of predicting the coarse-grained diffusivity has the
advantage that it constrains the diffusive fluxes in the coarse model to be
downgradient, unlike if we had predicted the tendency due to diffusion.
This approach also had the advantage that the same diffusivity is
applied to all thermodynamic and moisture variables, unlike if we had predicted the
effective diffusivity based on coarse-grained fluxes and gradients for each
variable separately.}
\textcolor{black}{The coarse-grained diffusivity is not predicted above $5.7$km, and the diffusivity from SAM at coarse resolution is used instead for these levels.}

Surface precipitation is not predicted separately by the RF parameterization but is rather diagnosed (including any surface sedimentation) as the sum of the 
resolved precipitation and the subgrid correction ($P_{\rm{tot}}^{\rm{subgrid}}(z = 0) + S^{\rm{subgrid}} (z = 0)$)
which is calculated from water conservation as
\begin{equation}
P^{\rm subgrid }_{\rm{tot}}(z = 0) + S^{\rm subgrid}(z = 0)= - \int_{0}^{\infty}  \left( q_{\rm{p}}^{\rm{subg-tend}}+q_{\rm{T}}^{\rm{subg-tend}} \right) \rho_{\rm{0}} dz .
\label{eq:subgrid_precipitation}
\end{equation}

\subsubsection*{Training and implementation \label{sub:Train / test}}
Before training the RFs, each output variable is standardized by removing the mean and rescaling to unit variance. For output variables with multiple vertical levels, the mean and variance are calculated across all levels used for that output variable.

We use $337.5$ days of 3-hourly model  output from the hi-res simulation to calculate the  features and outputs of the RFs. This model output was divided into a training dataset, validation dataset and a test dataset. The training dataset was obtained from the first $270$ days ($80\%$ of the data) of the hi-res simulation, the validation data set was obtained from the following  $33.75$ days ($10\%$ of the data), and the test data was obtained from the last $33.75$ days ($10\%$ of the data).  After tuning the hyperparameters, we expanded the training dataset to include the validation dataset for use in the final training process of the RFs used in SAM. 

To make the samples more independent, at each time step that was used, we randomly subsample atmospheric columns at each latitude. For coarse-graining factors of x4, x8 and x16,  we randomly select $10$, $20$ and  $25$  longitudes, respectively, at each latitude for every time step. For x32, the amount of coarse-grained output is relatively limited and so we do not subsample. This results in test and validation dataset sizes of $972,360$ samples for x4 and x8, $607,770$ samples for x16 and $218,790$ samples for x32. The amount of training data used is one of the hyperparameters we tuned as described below.

To train the RFs, we use the RandomForestRegressor class from scikit-learn 
package \cite{pedregosa2011scikit} version 0.21.2. 
Different hyperparameters governing the learning process and complexity of the RFs may be tuned to improve performance.
The most important hyperparameters that we tuned are the number of trees in each forest, the minimum number of samples at each leaf node, and the number of training samples. 
\textcolor{black}{Supplementary Figure 8}
shows the coefficient of determination ($R^2$) evaluated on the validation dataset for different combinations of hyperparameters. 
We stress that unlike standard supervised machine learning tasks, higher accuracy on test data is not our only goal. We also want to have a fast RF since it will be called many times when used in a simulation, and we do not want to have an RF that is overly large in memory since it will need to be stored on each core (or possibly shared across all cores in a node).
Based on a compromise between RF accuracy, memory demands and speed when the RF is implemented in SAM, for coarse-graining factors of x4, x8 and x16 we chose $10$ trees in each RF, a minimum of $20$ samples in each leaf and $5,000,000$ training samples.
However, fewer training samples were available for x32, and in order to have a similar size of RFs in this case, a minimum of $7$ samples in each leaf were taken.

\textcolor{black}{Training typically takes less than an hour using 10 CPU cores.}
For x8, RF-tend is  $0.75$GB and RF-diff is $0.20$GB when stored in netcdf
format at single precision.  We found that this size in memory did not pose a
problem when running across multiple cores.  We also emphasize that the RF
parameterization can achieve similar accuracy at a smaller size.  For example,
we reduced the number of trees in RF-tend from 10 to 5 which reduces its size
in memory by more than a factor of two to $0.35$GB without any noticeable
difference in the results when it is implemented in SAM at coarse resolution.
\textcolor{black}{Furthermore,  there are available techniques to reduce the
memory needed to store
RFs \cite{geurts2000some,bernard2009selection,painsky2018lossless} in case
memory becomes a limiting factor when 
using an RF parameterization in
operational climate simulations with more degrees of freedom.}
\textcolor{black}{Each RF was stored as a netcdf file, and routines to read in the netcdf files and to use the RFs to calculate outputs were added to SAM (using Fortran 90).}

\subsubsection*{Offline performance}
Offline performance is \textcolor{black}{primarily} evaluated using the coefficient of determination ($R^2$) as applied to the unscaled output variables  \textcolor{black}{in the test dataset}. $R^2$ is plotted for outputs of the RF parameterization as a function of the latitude and pressure in \textcolor{black}{Supplementary} Figure~9.  \textcolor{black}{For reference,} \textcolor{black}{the standard deviation of true outputs is plotted in \textcolor{black}{Supplementary} Figure~10} \textcolor{black}{and the mean of the true outputs is plotted in \textcolor{black}{Supplementary} Figure~1}. $R^2$ is generally higher in the lower and middle troposphere, though performance does vary across outputs.
Generally, the RFs tend to underestimate the variance in predictions compared to the true variance, although less so for larger coarse-graining factors (\textcolor{black}{Supplementary} Figure~11).
$R^2$ for the different outputs (combining data from all vertical levels for a given output) at different coarse-graining factors are given in \textcolor{black}{Supplementary} Table~2,  \textcolor{black}{and corresponding values of the root mean square error (RMSE) are given in \textcolor{black}{Supplementary} Table~3}.

RF-tend is also able to accurately predict the instantaneous surface precipitation rate (\textcolor{black}{Supplementary} Figure~3) with $R^2=0.99$ based on the test dataset for x8.  The predicted precipitation (including any surface sedimentation) is the sum of the resolved precipitation and the predicted subgrid correction ($P^{\rm subgrid}_{\rm{tot}}(z = 0) + S^{\rm subgrid}(z = 0)$) which is calculated from \textcolor{black}{equation 7}.

\textcolor{black}{
To further investigate the offline performance at different \textcolor{black}{grid spacing}, we focus on a comparison between x4 and x32. We coarse grain the subgrid tendencies calculated and predicted at x4 to the same grid as x32 (referred to as x4$\rightarrow$x32) such that they are on the same grid as the subgrid tendencies calculated and predicted at x32. 
To do the coarse graining, it was necessary to make an alternative  test dataset since the default test dataset is randomly subsampled in longitude for x4. $100$ snapshots from the hi-res simulation were used without subsampling in longitude. 
This results in an alternative test dataset size of $5,184,000$ for the x4 case and $81,000$ for the x4$\rightarrow$x32 and x32 cases. \textcolor{black}{We find that calculating  $R^2$ and RMSE from these $100$ snapshots gives almost identical results compared to the test dataset that was used for model evaluation (compare \textcolor{black}{Supplementary} Tables~2 and 3 to 4)}.} One snapshot from the alternative test dataset is shown in \textcolor{black}{Fig. 4}, and we also use the alternative test dataset for the results shown in \textcolor{black}{Supplementary} Figure~5 \textcolor{black}{and in \textcolor{black}{Supplementary} Table~4.}

\clearpage
\begin{itemize}
\item [Acknowledgements] We thank Bill Boos for providing the output from the high-resolution simulation, and we thank Daniel Koll, Nick Lutsko \textcolor{black}{and Chris Hill} for helpful discussions.  
We acknowledge high-performance computing support from Cheyenne (doi:10.5065/D6RX99HX) provided by NCAR's Computational and Information Systems Laboratory, sponsored by the National Science Foundation.
We acknowledge support from the MIT Environmental Solutions Initiative, the EAPS Houghton-Lorenz postdoctoral fellowship, and NSF AGS-1552195.

 \item[Competing Interests] The authors declares that they have no
competing interests.
\item[Author contribution] J.Y. and P.O. designed the research. 
J.Y. performed numerical simulations. 
J.Y.  and P.O.  wrote the manuscript.
 \item[Correspondence] Correspondence and requests for materials
should be addressed to J.Y.~(email: janniy@mit.edu).
\end{itemize}


\clearpage
\bibliography{yanibib}

\end{document}


\date{}
\maketitle

\renewcommand{\thepage}{S\arabic{page}} 
\renewcommand{\thesection}{S\arabic{section}}  
\renewcommand{\thetable}{S\arabic{table}}  
\renewcommand{\thefigure}{S\arabic{figure}}
\renewcommand{\theequation}{S\arabic{equation}}

\clearpage
\section*{Supplementary Note 1\label{sub:Conservation of energy}}
\begin{figure}
\begin{centering}
\includegraphics[scale=0.8]{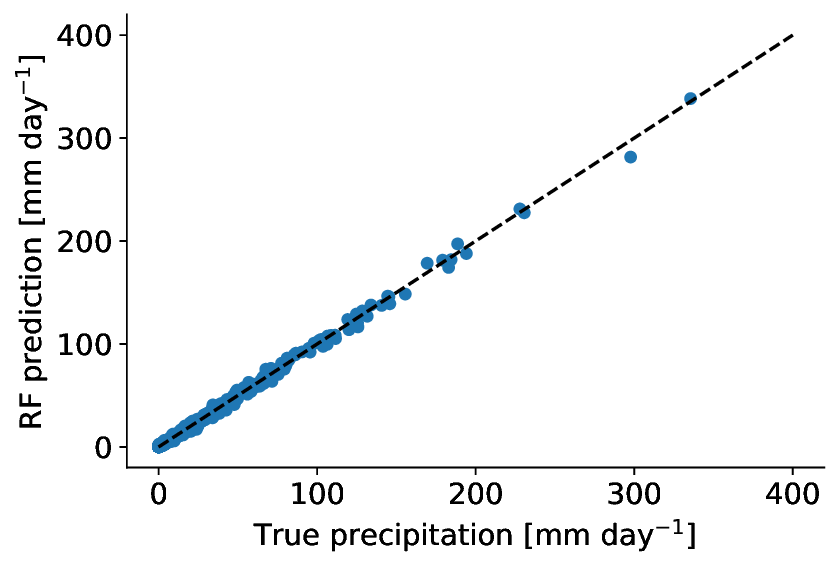} 
\par\end{centering}
\protect\caption{
\textbf{\textcolor{black}{Offline performance for surface precipitation.}}
Scatter plot of true instantaneous surface precipitation coarse-grained to x8 versus the \textcolor{black}{random forest (RF)} prediction. The RF-predicted precipitation is calculated as the sum of the resolved precipitation and the subgrid correction.  A random subset of 10,000 samples from the test set are shown for clarity. The black dashed line is the one-to-one line. We verified that the RF prediction gives non-negative precipitation values for all the $972,360$ test samples. 
\label{fig:Precip_scatter_RF_vs_True}}
\end{figure}

\textcolor{black}{Here we show that the RF parameterization performs well in respecting physical constraints.}
\textcolor{black}{This good performance arises}
because the constraints are respected by the training data and the RF
predictions are averages over subsets of the training
data\cite{Ogorman2018using}.  In particular, the RF parameterization always
predicts non-negative surface precipitation
(\textcolor{black}{Supplementary} Figure~\ref{fig:Precip_scatter_RF_vs_True}).  Similarly, in the remainder of this
section we show that the RF parameterization conserves energy in the absence of
external forcing (i.e., in the absence of radiative heating and surface fluxes of $h_{\rm{L}}$). 

RF-diff automatically respects energy conservation in the absence of external forcing since it predicts the turbulent diffusivity rather than the diffusive tendencies.  To check energy conservation for RF-tend, we integrate the evolution equation for $h_{\rm{L}}$  (\textcolor{black}{equation 1 in the methods section})
 in  the vertical with density weighting, and then consider the contributions to the resulting equation that come from RF-tend (denoted with a superscript subgrid) to give an energy-conservation residual:  
\begin{equation}
{\rm residual} = \int_{0}^{\infty} \rho_{\rm{0}} \left(\frac{\partial h_{\rm{L}}}{\partial t}\right)^{\rm subgrid}_{\rm no-rad} dz + \overline{L}_p P_{\rm tot}^{\rm subgrid}(z = 0) + \overline{L}_n S^{\rm\, subgrid}(z=0).
\label{eq:residuals}
\end{equation}
Here $\left(\frac{\partial h_{\rm{L}}}{\partial t}\right)^{\rm subgrid}_{\rm no-rad}$ is the subgrid tendency of $h_{\rm{L}}$ but excluding the contribution from radiative heating which is an external forcing. This tendency was evaluated by training a new RF-tend that predicts the radiative heating tendency and the sum of other tendencies of $h_{\rm{L}}$ as separate outputs. This RF-tend performed similarly to our default RF-tend in all other regards.
In deriving equation~\ref{eq:residuals}, we have neglected subgrid correlations between $L_{\rm{p}}$ and $P_{\rm tot}$ and between $L_n$ and $S$, and as a result the residual will not be exactly zero even for the true subgrid tendencies. In addition, in evaluating the residual, the column energy change due to subgrid surface precipitation and sedimentation ($\overline{L}_p P_{\rm tot}^{\rm subgrid}(z = 0) + \overline{L}_n S^{\rm\, subgrid}(z=0)$) was approximated  to be $\overline{L}_p (P^{\rm{subgrid}}_{\rm{tot}}(z=0) +  S^{\rm{subgrid}}(z=0))$ so that we could evaluate it using equation \textcolor{black}{7 in the methods section}.
This approximation leads to a small error to the extent that there is surface sedimentation.

\begin{figure}
\begin{centering}
\includegraphics[scale=0.85]{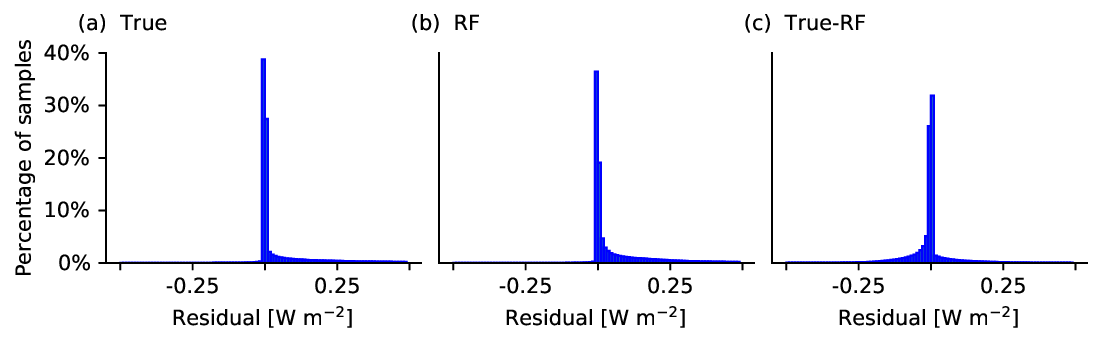} 
\par\end{centering}
\protect\caption{
\textbf{\textcolor{black}{Distribution of energy-conservation residuals.}}
\textcolor{black}{Equation \ref{eq:residuals} is} applied to samples in the test dataset at x8 for the (a) true subgrid tendencies, (b) subgrid tendencies \textcolor{black}{predicted by the random forest parameterization}, and (c) the difference between the true and predicted subgrid tendencies. The bin size is $0.01 \rm{W m^{-2}}$. \label{fig:energy_conservation}}
\end{figure}

The distribution of the energy-conservation residual for the true subgrid tendencies is shown in \textcolor{black}{Supplementary} Figure~\ref{fig:energy_conservation}a and for the RF-predicted subgrid tendencies in \textcolor{black}{Supplementary} Figure~\ref{fig:energy_conservation}b.  In general, the residuals are very small, and the distribution of the residuals is similar for the true subgrid tendencies and the RF-predicted subgrid tendencies. The difference between the true and the RF-predicted residuals for each column was also calculated, and its distribution is shown in \textcolor{black}{Supplementary} Figure~\ref{fig:energy_conservation}c. 

\textcolor{black}{Supplementary} Figure~\ref{fig:energy_conservation}b demonstrates that the RF parameterization respects energy conservation to a high degree of accuracy (less than $2\%$ of the data has residuals that are larger in amplitude than $1 \rm{W~m^{-2}}$). The root mean square error in energy conservation is
 $0.35\rm{W~m^{-2}}$ \textcolor{black}{which is much smaller than the root mean square value of $64.76\rm{W~m^{-2}}$ for the vertical integral of the predicted energy tendencies}. The mean bias error is $0.11\rm{W~m^{-2}}$.  We note that a similar mean bias error is found in the calculation of the energy-conservation residual from the true subgrid tendencies, and both  are likely a result of the approximations we used in the calculation of the energy conservation residual rather than a violation of energy conservation
(the mean bias error found in \textcolor{black}{Supplementary} Figure~\ref{fig:energy_conservation}c is $0.0001\rm{W~m^{-2}}$).
The root mean square error of $0.35\rm{W~m^{-2}}$ is substantially smaller than a reported value of $92\rm{W~m^{-2}}$ in a previous study that used a NN to learn from a quasi-global simulation\cite{brenowitz2018prognostic} with the caveat that the metric of errors in energy conservation in that study also included errors in predicted radiative heating and surface fluxes. We note also that energy conservation for a NN parameterization can be enforced by including it as a constraint in the NN architecture\cite{beucler2019achieving}. 

\clearpage
\section*{Supplementary Note 2 \label{SI:RF without qp}}
\begin{figure}
\begin{centering}
\includegraphics[scale=1.0]{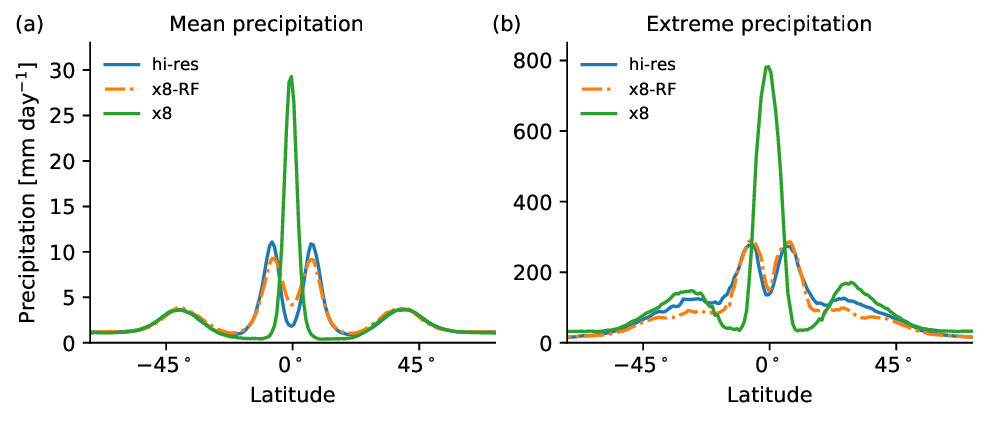} 
\par\end{centering}
\protect\caption{
\textbf{\textcolor{black}{Mean and extreme precipitation in simulation with the alternative random forest-parameterization.}}
\textcolor{black}{The simulation with the alternative random forest parameterization (Supplementary Note 2) does not use the precipitating water \textcolor{black}{($q_p$)} as a variable.} \textcolor{black}{Shown are} (a) mean precipitation and (b) 99.9th percentile of 3-hour precipitation at each latitude from the hi-res simulation (blue), x8-RF simulation without $q_{\rm{p}}$ (orange), and x8 simulation (green).} \label{fig:Online_x8_precip_NO_QP}
\end{figure}

Here we describe an alternative RF parameterization approach in which $q_{\rm{p}}$ is not used as a variable. This alternative RF parameterization leads to stable simulations when implemented in SAM, and it gives similarly accurate results to the default approach for mean precipitation, but less accurate results for extreme precipitation in midlatitudes (\textcolor{black}{Supplementary} Figure~\ref{fig:Online_x8_precip_NO_QP}). Since the alternative RF parameterization does not take $q_{\rm{p}}$ as an input, SAM in this case does not include $q_{\rm{p}}$ as a prognostic variable.  Such a parameterization could be potentially very useful since $q_{\rm{p}}$ is a variable \textcolor{black}{that changes on short time scales and therefore} limits the size of the time step at coarse resolution.  \textcolor{black}{Furthermore, many climate models do not use $q_{\rm{p}}$ as a prognostic variable, and using an ML parameterization in these models requires a parameterization that does not use $q_{\rm{p}}$ as an input.
Note that we haven't yet tested this parameterization with much larger time steps because turbulent diffusion as implemented in SAM also limits the time step. }

The equations for the prognostic water and energy variables in SAM are described in \textcolor{black}{equations 1-3 in the methods section}.
We define a new prognostic energy variable ($H_{\rm{L}}$) that does not include the precipitating water ($q_{\rm{p}}$):
\begin{equation}
H_{\rm{L}} = c_{\rm{p}}T + gz - L_{\rm{c}} q_c - L_{\rm{s}} q_{\rm{i}}.
\end{equation}
This can be written in terms of the original energy variable $h_{\rm{L}}$ as  
\begin{equation}
H_{\rm{L}} = h_{\rm{L}} + L_{\rm{p}} q_{\rm{p}}
\label{eq:HL_definition}
\end{equation}
where, $L_{\rm{p}} = L_{\rm{c}} + L_{\rm{f}}(1-\omega_{\rm{p}})$ and $\omega_{\rm{p}}$ is the partition function for precipitation which depends only on temperature in SAM\cite{khairoutdinov2003cloud}. In the following, we account for vertical variations of $L_{\rm{p}}$ in the vertical but neglect the smaller variations in the horizontal and in time.
Taking the derivative with respect to time of equation~\ref{eq:HL_definition} gives
 \begin{equation}
 \frac{\partial H_{\rm{L}}}{\partial t} =  \frac{\partial h_{\rm{L}}}{\partial t}  + L_{\rm{p}}  \frac{\partial q_{\rm{p}}}{\partial t}.
\label{eq:HL prognostic idea}
\end{equation}
Substituting equations \textcolor{black}{1 and 3 from the methods section}
into equation~\ref{eq:HL prognostic idea}, we get a prognostic equation for $H_{\rm{L}}$:
\begin{align}
\frac{\partial H_{\rm{L}}}{\partial t} &= -\frac{1}{\rho_{\rm{0}}} \frac{\partial}{\partial x_i}(\rho_{\rm{0}} u_i H_{\rm{L}} ) 
-\frac{1}{\rho_{\rm{0}}}  \frac{\partial}{\partial z}(L_n S)
+L_{\rm{p}}\left(\frac{\partial q_{\rm{p}}}{\partial t}\right)_{\rm{mic}} + \left(\frac{\partial h_{\rm{L}}}{\partial t}\right)_{\rm{rad}}   \nonumber \\
&  - \frac{1}{\rho_{\rm{0}}} \frac{\partial F_{H_{\rm{L}} i}}{\partial x_i}
 + \frac{1}{\rho_{\rm{0}}} \frac{\partial L_{\rm{p}}}{\partial z} (\rho_{\rm{0}} w q_{\rm{p}} + F_{q_{\rm{p}} z}  -  P_{\rm{tot}}) \label{eq_HL_tend}
\end{align}
where $F_{H_{\rm{L}} i} =F_{h_{\rm{L}} i} +  L_{\rm{p}} F_{q_{\rm{p}} i}$ and the last term on the right hand side results from heating from phase changes of precipitation.

Our aim is to make a parameterization for coarse-resolution simulations that does not include $q_{\rm{p}}$.  Therefore we assume that at coarse resolution we can neglect  the horizontal fluxes of $q_{\rm{p}}$ and the time derivative of $q_{\rm{p}}$ in equation~\textcolor{black}{3 from the methods section}.
 Integrating \textcolor{black}{this equation}
vertically over the column and neglecting surface diffusive fluxes of $q_{\rm{p}}$ then gives an expression for the surface precipitation rate:
\begin{equation}
P_{\rm{tot}}(z=0) = -\int_{0}^{\infty} \left(\frac{\partial q_{\rm{p}}}{\partial t}\right)_{\rm{mic}}  dz. 
\label{eq:precip_noqp}
\end{equation}

The RF parameterization without $q_{\rm{p}}$ is similar in most respects to the RF parameterization with $q_{\rm{p}}$, but some changes are needed.
First, RF-tend does not use $q_{\rm{p}}$ as a feature or predict its tendency as an output, and it predicts the tendency of $H_{\rm{L}}$ rather than the tendency of $h_{\rm{L}}$.
Thus, the features for RF-tend are $X = (T ,q_{\rm{T}},|y|)$, and the outputs are $y = (H_{\rm{L}}^{\rm{subg-tend}},q_{\rm{T}}^{\rm{subg-tend}})$. RF-diff is changed to predict the subgrid surface flux of $H_{\rm{L}}$ instead of $h_{\rm{L}}$.
Second, RF-tend in this version predicts for $H_{\rm{L}}^{\rm{subg-tend}}$  the subgrid vertical advection and subgrid sedimentation terms added to the total value of
$L_{\rm{p}}\left(\frac{\partial q_{\rm{p}}}{\partial t}\right)_{\rm{mic}} + 
\left(\frac{\partial h_{\rm{L}}}{\partial t}\right)_{\rm{rad}} +
\frac{1}{\rho_{\rm{0}}} \frac{\partial L_{\rm{p}}}{\partial z} (\rho_{\rm{0}} w q_{\rm{p}} + F_{q_{\rm{p}} z}  -  P_{\rm{tot}})$ in equation \ref{eq_HL_tend}.
Third, we do not apply the RF tendency of $q_{\rm{T}}$ due to subgrid vertical advection and sedimentation above $11.8\rm{km}$ to avoid a feedback that lead to a severe change in the global circulation. 
(This is likely to be a similar issue to an instability that occurred in a previous study on ML parameterization that also did not use $q_{\rm{p}}$ as a prognotic variable and in which this instability was dealt with by not including certain upper-level variables as features\cite{brenowitz2019spatially}.)
To avoid over-fitting the results presented here, we chose the same upper-level cutoff for these $q_{\rm{T}}$ tendencies ($11.8\rm{km}$) as was also used for radiative heating. We tested different upper-level cutoffs
($11\rm{km}$, $9.5\rm{km}$) and different combinations of cutoff levels (different cutoff levels for each process) and found that all these choices led to simulations with qualitatively similar results.

When implementing the alternative RF parameterization in SAM, we remove $q_{\rm{p}}$ as a prognostic variable and change from $h_{\rm{L}}$ to $H_{\rm{L}}$ as a prognostic variable. We diagnose surface precipitation using equation \ref{eq:precip_noqp} (plus any surface sedimentation).  The approximations used in deriving equation \ref{eq:precip_noqp} can result in negative instantaneous surface precipitation in rare cases. However, the surface precipitation averaged over $3$ hours in the SAM simulations with this RF parameterization is negative less than 1\% of the time and the negative values are smaller in magnitude than $0.2\rm{mm~day^{-1}}$. 

\clearpage
\begin{figure}
\begin{centering}
\includegraphics[scale=1.1]{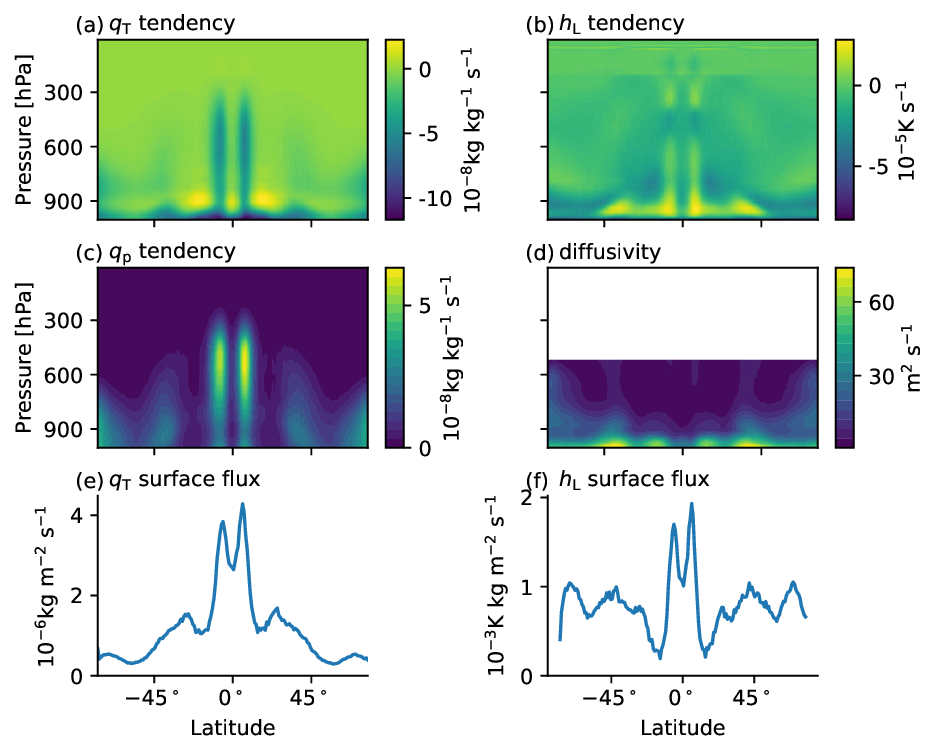} 
\par\end{centering}
\protect\caption{
\textbf{\textcolor{black}{Mean true outputs of random forest parameterization.}}
\textcolor{black}{The time- and zonal-mean for different true outputs of the \textcolor{black}{random forest parameterization} at x8: 
(a) subgrid tendency of $q_{\rm{T}}$, 
(b) subgrid tendency of $h_{\rm{L}}$, 
(c) subgrid tendency of $q_{\rm{p}}$, 
(d) $\overline{D}$,
(e) subgrid surface flux of $q_{\rm{T}}$, and 
(f) subgrid surface flux of $h_{\rm{L}}$.}
 \label{fig:Offline_mean}}
\end{figure}

\clearpage

\clearpage

\clearpage

\clearpage

\thispagestyle{empty}
\begin{figure}
\begin{centering}
\includegraphics[scale=0.85]{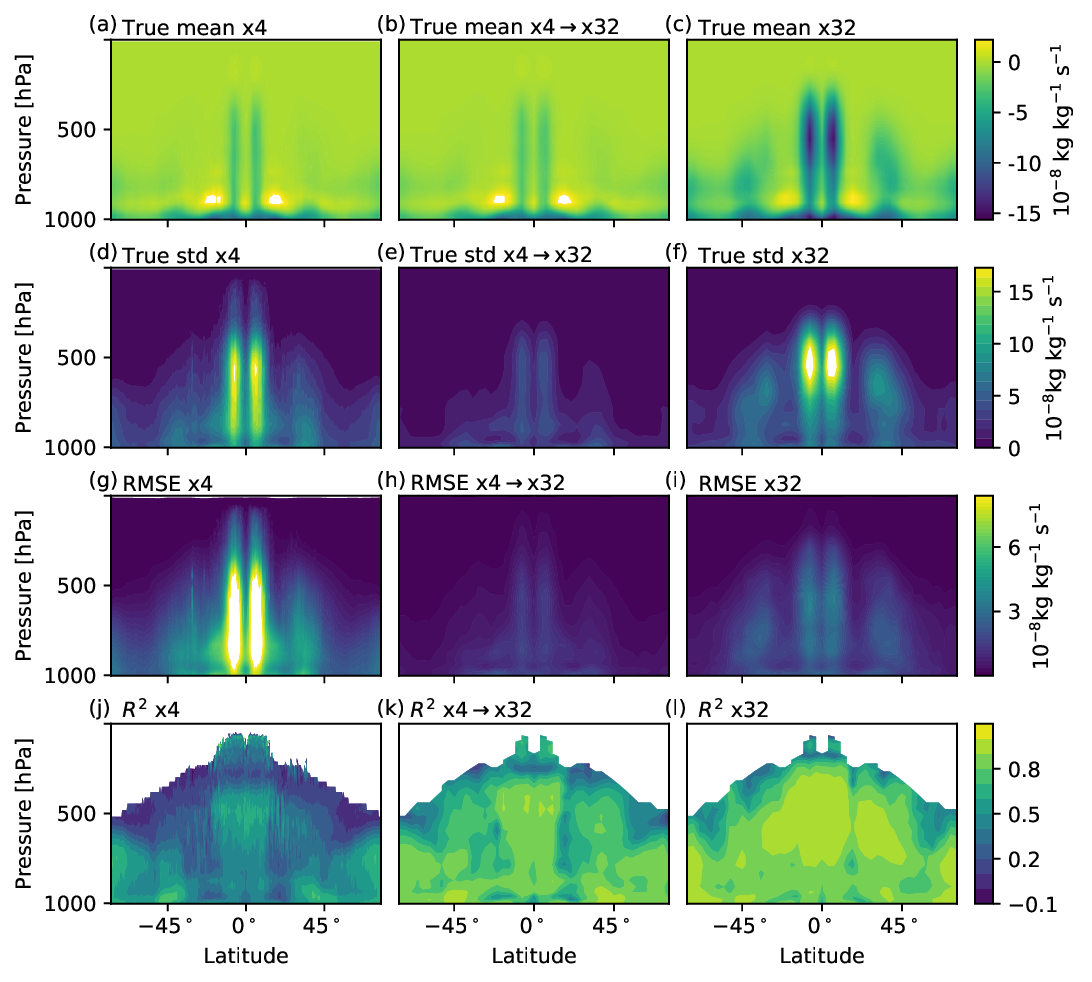} 
\par\end{centering}
\protect\caption{
\textcolor{black}{
\textbf{\textcolor{black}{Offline comparison of parameterizations at a common grid spacing.}}
Offline results for the subgrid tendency of $q_{\rm{T}}$:
(a-c) true mean, 
(d-f) true standard deviation,  
(g-i) \textcolor{black}{root mean square error} for the  \textcolor{black}{random forest (RF)} prediction,  and
(j-l) coefficient of determination ($R^2$) for the RF prediction.
Results are shown for x4 (\textcolor{black}{a,d,g,j}), the subgrid tendency calculated and predicted at x4 and then coarse-grained to x32 (\textcolor{black}{b,e,h,k}), and x32 (\textcolor{black}{c,f,i,l}). 
\textcolor{black}{Results shown in this figure are based on the alternative test dataset
(see \textcolor{black}{methods}).
Colorbar is saturated in panels f and g.}
 }} \label{fig:qt_rmse_mean_std_R2_x4_to_x32}
\end{figure}
\clearpage

\thispagestyle{empty}
\begin{figure}
\begin{centering}
\includegraphics[scale=1.0]{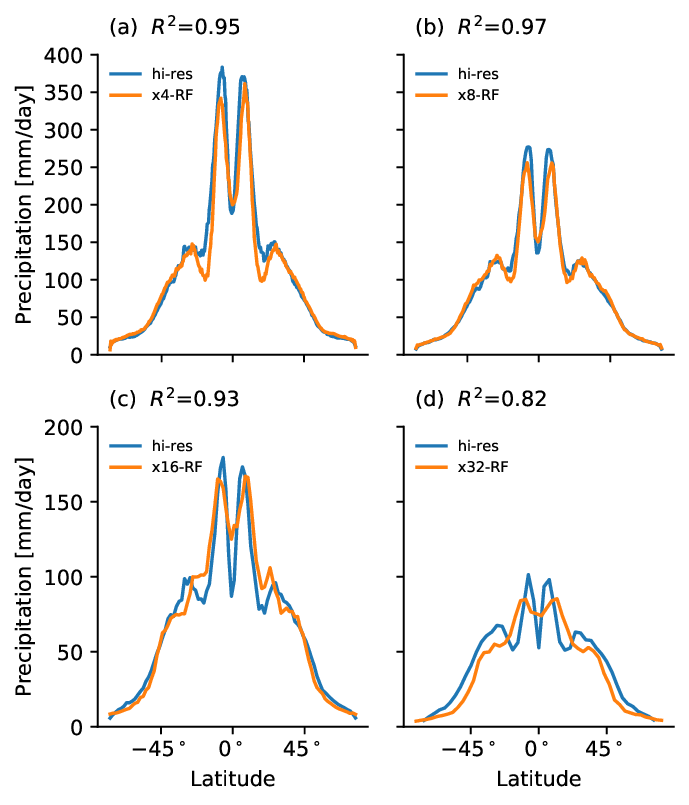} 
\par\end{centering}
\protect\caption{
\textbf{\textcolor{black}{Extreme precipitation for simulations with different horizontal grid spacing.}}
Extreme precipitation as a function of latitude as measured by the $99.9$th percentile of 3-hourly precipitation \textcolor{black}{for:} (a) x4, (b) x8, (c) x16 and (d) x32. \textcolor{black}{Results are shown} for the \textcolor{black}{hi-res} simulation (blue) and the coarse-resolution simulation with the \textcolor{black}{random forest} parameterization (orange). The precipitation rates for the \textcolor{black}{hi-res} simulation have been coarse-grained to the appropriate grid spacing prior to calculating the percentiles\cite{chen2008verification}. \textcolor{black}{The coefficient of determination for each of the grid spacings is given above each panel.}} \label{fig:extreme_precip_x8_x16_x32_qp_Latin}
\end{figure}

\clearpage

\thispagestyle{empty}
\begin{figure}
\begin{centering}
\includegraphics[scale=0.60]{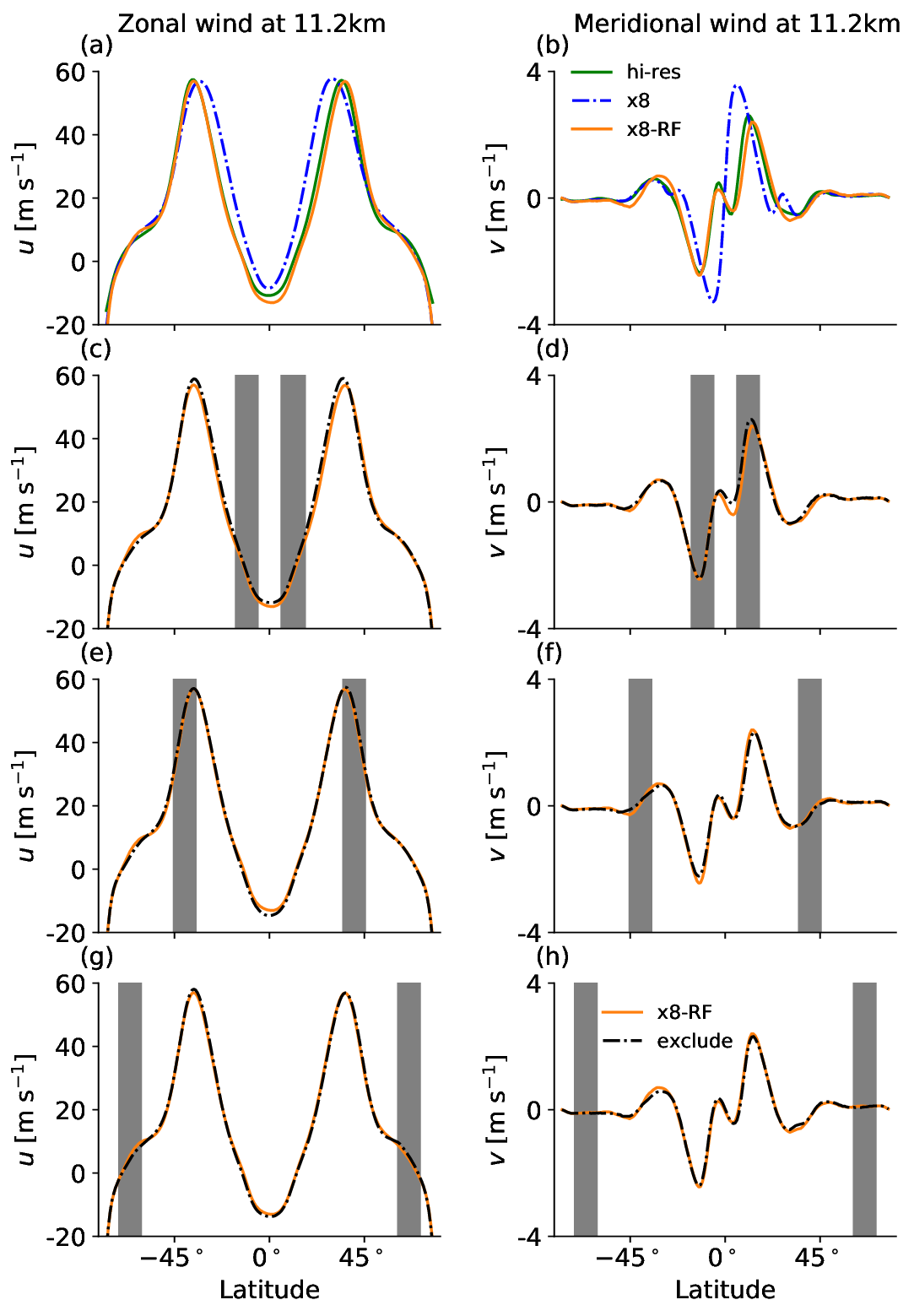} 
\par\end{centering}
\protect\caption{
\textbf{\textcolor{black}{Online performance when excluding latitude bands during the training process.}}
\textcolor{black}{
Zonal- and time-mean of (a, c, e, g) the zonal wind at $11.2$km  and (b, d,  f, h) 
the meridional wind at $11.2$km. 
(a,b) The hi-res (green) and x8  (blue dash-dotted) \textcolor{black}{simulations.}
(c-h) \textcolor{black}{Simulations} with x8-RFs \textcolor{black}{(black dash-dotted)} in which the training process excludes 
in both hemispheres the latitude bands 
(c,d) $5.1^{\circ}-15.5^{\circ}$,
(e,f) $34.5^{\circ}-44.9^{\circ}$ and 
(g,h) $60.5^{\circ}-70.8^{\circ}$.
For comparison the results for x8-RF without any latitudes excluded in training are plotted in orange in all panels. 
Grey bars indicate latitude bands that where excluded during the training.}} \label{fig:U_EKE_Exclude_6_18_40_52_70_82_offline_online_vert_two_RFs}
\end{figure}
\clearpage

\begin{figure}
\begin{centering}
\includegraphics[scale=0.8]{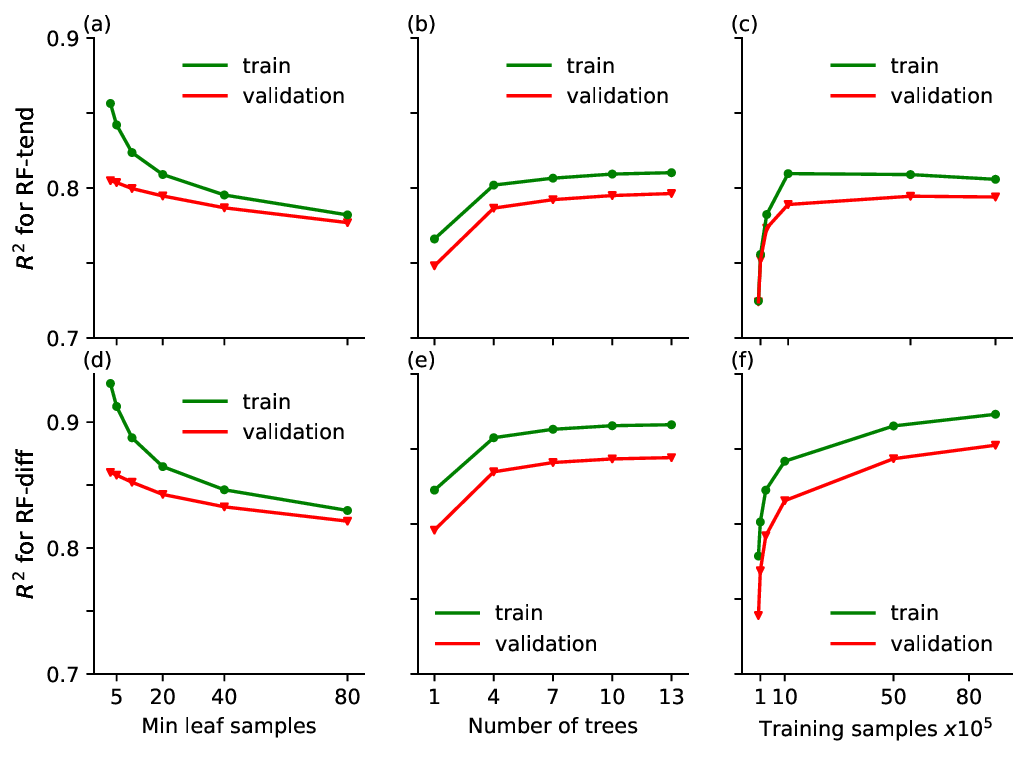} 
\par\end{centering}
\protect\caption{
\textbf{\textcolor{black}{Hyperparameter tuning for random forests.}}
Coefficient of determination ($R^2$) for RF-tend (panels a-c, $R^2$ calculated for $q_{\rm{T}}^{\rm{subg-tend}}$) and RF-diff (d-f, $R^2$ calculated for diffusivity) as evaluated on the training dataset (green) and validation dataset (red) for x8 and different hyperparameter values:  (a,d) minimum samples in each leaf, (b,e) number of trees in the forest, and (c,f)  number of training samples.  The hyperparameters that are used for both \textcolor{black}{random forests (RFs)} when implemented in SAM are $20$ minimum samples in each leaf, $10$ trees in the forest and $5,000,000$ training samples for the x4, x8, and x16 simulations. The hyperparameters that are used for both RFs in the x32 simulations are $7$ minimum samples in each leaf, $10$ trees in the forest and $1,969,020$ training samples.
 \label{fig:hyperparameter_tuning}}
\end{figure}

\clearpage

\thispagestyle{empty}
\begin{figure}
\begin{centering}
\includegraphics[scale=1.1]{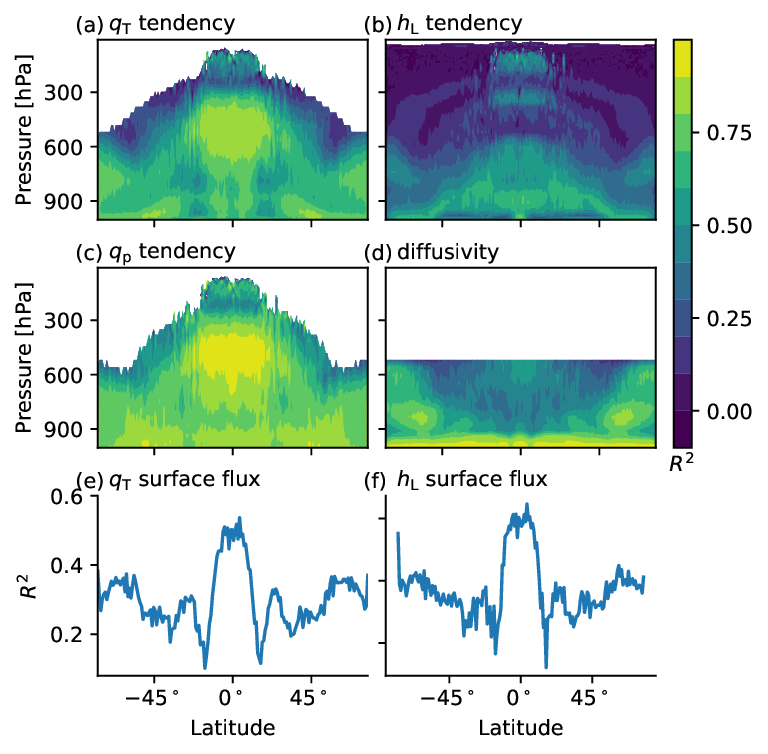} 
\par\end{centering}
\protect\caption{
\textbf{\textcolor{black}{Offline performance of random forest parameterization.}}
 \textcolor{black}{The offline performance measured by the coefficient of determination ($R^2$)} at x8 for: 
(a) subgrid tendency of $q_{\rm{T}}$, 
(b) subgrid tendency of $h_{\rm{L}}$, 
(c) subgrid tendency of $q_{\rm{p}}$, 
(d) $\overline{D}$,
(e) subgrid surface flux of $q_{\rm{T}}$, and 
(f) subgrid surface flux of $h_{\rm{L}}$.
Results are based on the samples from the test dataset. $R^2$ is only shown where the variance is at least $0.1$\% of the mean variance over all latitudes and levels.   \label{fig:Offline_Rsq}}
\end{figure}

\clearpage

\begin{figure}
\begin{centering}
\includegraphics[scale=1.1]{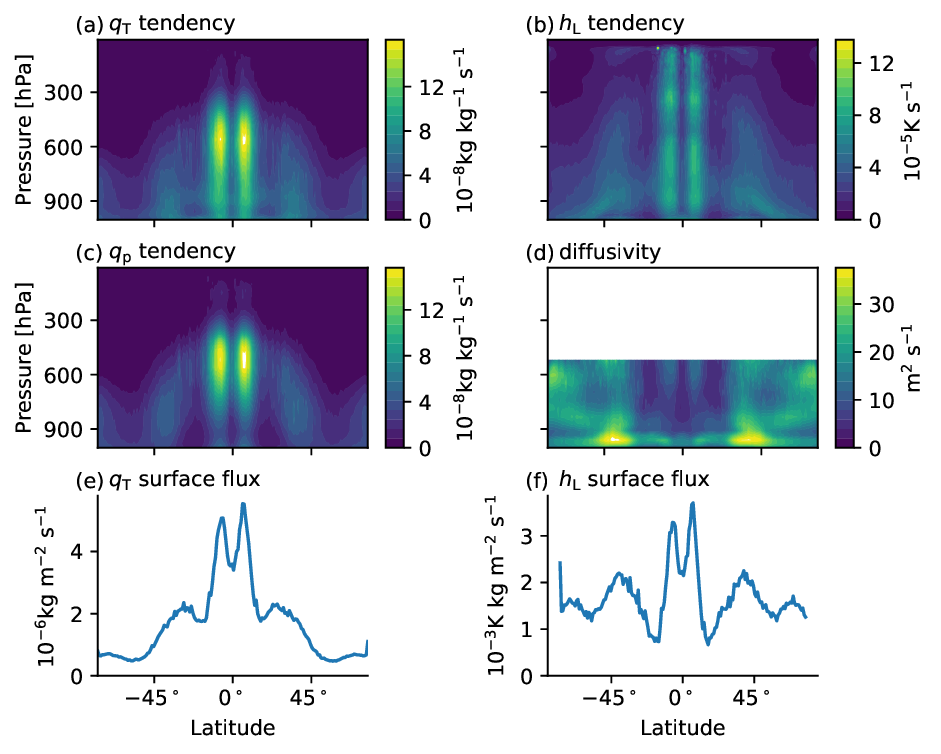} 
\par\end{centering}
\protect\caption{
\textbf{\textcolor{black}{Standard deviation of true outputs for random forest parameterization.}}
\textcolor{black}{
The standard deviation of true outputs at x8:
(a) subgrid tendency of $q_{\rm{T}}$, 
(b) subgrid tendency of $h_{\rm{L}}$, 
(c) subgrid tendency of $q_{\rm{p}}$, 
(d) $\overline{D}$,
(e) subgrid surface flux of $q_{\rm{T}}$, and 
(f) subgrid surface flux of $h_{\rm{L}}$.}
 \label{fig:Offline_std}}
\end{figure}

\clearpage

\begin{figure}
\begin{centering} 
\includegraphics[scale=1.0]{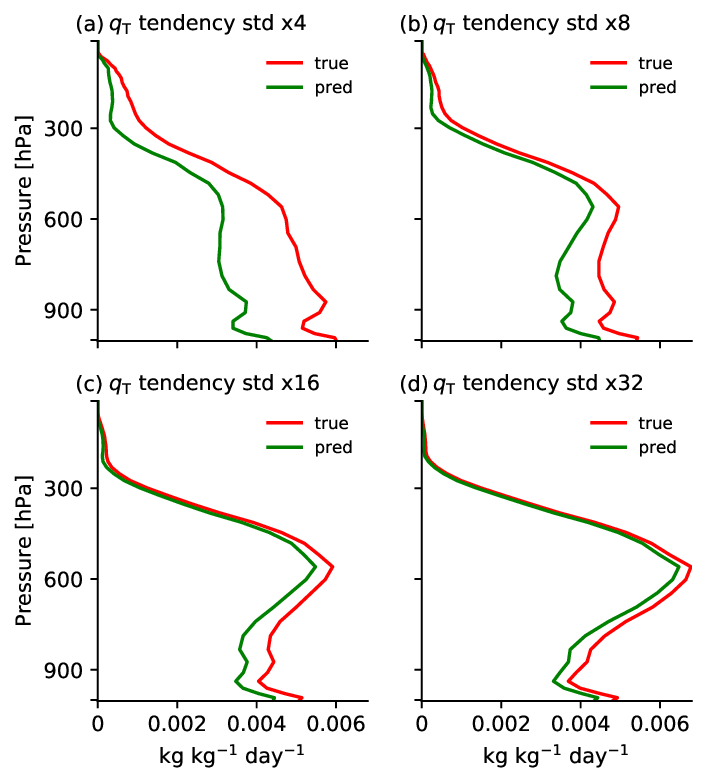} 
\par\end{centering}
\protect\caption{
\textbf{\textcolor{black}{Vertical profiles of the standard deviation of outputs for different coarse graining factors.}}
\textcolor{black}{The standard deviation of} true (red) and \textcolor{black}{random forest}-predicted  (green) subgrid tendency of $q_{\rm{T}}$ for different coarse-graining factors: (a) x4, (b) x8, (c) x16, and (d) x32. Results are evaluated based on the test dataset.} \label{fig:std_vs_pressure_qp_latin}
\end{figure}

\clearpage


\begin{table}[]
\centering
\begin{tabular}{lcc}
\hline
                    & \multicolumn{1}{l}{x8-RF} & x8 \\ \hline
Eddy kinetic energy ($\rm{m^2~s^{-2}}$) & 26.3 (0.97)                            & 54.5 (0.88)          \\
Zonal wind   ($\rm{m~s^{-1}}$)       & 2.2 (0.98)                         &  4.5 (0.87)      \\
Meridional wind ($\rm{m~s^{-1}}$)    & 0.2  (0.87)                         & 0.6 (-0.01)    \\
Non-precipitating water  ($\rm{g~kg^{-1}}$) & 0.1 (0.99)               & 0.4 (0.97)   \\ \hline
\end{tabular}
\caption{
\textbf{\textcolor{black}{Online performance with and without \textcolor{black}{random forest} parameterization.}}
Online performance as measured by  \textcolor{black}{root mean square error} ($R^2$ in parenthesis) of zonal- and time-mean variables for the coarse-resolution simulations with the  \textcolor{black}{random forest} parameterization (x8-RF) and without the \textcolor{black}{random forest} parameterization (x8) as compared to the target hi-res simulation.  The eddy kinetic energy is defined with respect to the zonal and time mean. }
\label{table:Rsq_yz_online_eke_u_v_qv}
\end{table}

\begin{table}[]
\centering
\begin{tabular}{lcccccc}
\hline
       & $q_{\rm{T}}^{\rm{subg-tend}}$ & $h_{\rm{L}}^{\rm{subg-tend}}$ & $q_{\rm{p}}^{\rm{subg-tend}}$ & $\overline{D}$ & $q_{\rm{T}}^{\rm{surf-flux}}$ & $h_{\rm{L}}^{\rm{surf-flux}}$ \\ \hline
x4  & 0.56              & 0.31              & 0.73              & 0.72        & 0.30               & 0.29              \\
x8  & 0.80              & 0.48              & 0.88              & 0.84        & 0.48              & 0.44              \\
x8-no-$|y|$  & 0.79              & 0.48              & 0.88              & 0.84        & 0.47              & 0.42              \\
x16 & 0.90              & 0.64              & 0.93              & 0.93        & 0.60              & 0.57             \\
x32 & 0.95              & 0.75              & 0.96              & 0.96        & 0.78              & 0.76             \\ \hline
\end{tabular}
\caption{ 
\textbf{\textcolor{black}{Offline performance of random forest parameterizations as measured by $\boldsymbol{R^2}$.}} \textcolor{black}{The offline performance \textcolor{black}{is given} for different coarse-graining factors and different outputs of the \textcolor{black}{random forests}.}
 \textcolor{black}{ For x8-no-$|y|$, the distance from equator was not used as a feature}. For the tendencies and turbulent diffusivity, all levels used are included when calculating $R^2$. All results are based on the test dataset.}
\label{table:offline performance}
\end{table}

\clearpage

\begin{table}[]
\centering
\begin{tabular}{lcccccc}
\hline

       & $q_{\rm{T}}^{\rm{subg-tend}}$ & $h_{\rm{L}}^{\rm{subg-tend}}$ & $q_{\rm{p}}^{\rm{subg-tend}}$ & $\overline{D}$ & $q_{\rm{T}}^{\rm{surf-flux}}$ & $h_{\rm{L}}^{\rm{surf-flux}}$ \\
& $10^{-8}\times$ & $10^{-5}\times$ & $10^{-8}\times$ &  & $10^{-6}\times$ & $10^{-3}\times$ \\      
&    kg kg$^{-1}$s$^{-1}$ &  K s$^{-1}$ &  kg kg$^{-1}$ s$^{-1}$ & m$^2$ s$^{-1}$ & kg m$^{-2}$ s$^{-1}$ &  K kg m$^{-2}$ s$^{-1}$ \\ \hline
x4  						& 2.50             &   3.65           & 1.23              & 15.27        & 1.76               & 1.36              \\
x8  						& 1.64            & 2.27              & 0.94              & 9.88        & 1.68              & 1.35              \\
x8-no-$|y|$  & 1.65             & 2.28             & 0.94             & 10.02        & 1.69              & 1.37             \\
x16 					& 1.22              & 1.54              & 0.83              & 6.64       & 1.76              & 1.52             \\
x32 					& 0.94              & 1.10              & 0.72              & 4.28        & 1.69              & 1.58             \\ \hline
\end{tabular}
\caption{\textcolor{black}{ 
\textbf{\textcolor{black}{Offline performance of random forest parameterizations as measured by root mean square error}}. \textcolor{black}{The offline performance \textcolor{black}{is given} for different coarse-graining factors and different outputs of the \textcolor{black}{random forests}.}
 \textcolor{black}{ For x8-no-$|y|$, the distance from equator was not used as a feature}. For the tendencies and turbulent diffusivity, all levels used are included when calculating \textcolor{black}{root mean square error}. All results are based on the test dataset.}}
\label{table:offline performance std}
\end{table}
\clearpage

\clearpage

\begin{table}[]
\centering
\begin{tabular}{lccc}
\hline

       & $q_{\rm{T}}^{\rm{subg-tend}}$ & $h_{\rm{L}}^{\rm{subg-tend}}$ & $q_{\rm{p}}^{\rm{subg-tend}}$  \\
&    $10^{-8}$ kg kg$^{-1}$s$^{-1}$ & $10^{-5}$ K s$^{-1}$ & $10^{-8}$ kg kg$^{-1}$ s$^{-1}$  \\ \hline
x4  							& 2.57 (0.55)           &   3.79 (0.29)          & 1.28    (0.71)                     \\
x4$\rightarrow$x32 & 0.49  (0.93)            & 0.39  (0.82)            & 0.21 (0.95)                          \\
x32 							& 0.97  (0.95)            & 1.16   (0.73)           & 0.74 (0.96)                          \\ \hline
\end{tabular}
\caption{\textcolor{black}{
\textbf{\textcolor{black}{Offline comparison of random forest parameterizations at a common grid spacing.}}
Offline performance as measured by \textcolor{black}{root mean square error}  ($R^2$ values in brackets) for different outputs of RF-tend \textcolor{black}{for x4,  
coarse graining of the subgrid tendencies calculated and predicted at x4
to x32 grid spacing (x4$\rightarrow$x32), and x32. Results in this table are based on the alternative test dataset (see \textcolor{black}{methods}).}}}
\label{table:offline performance x4 to x32}
\end{table}

\clearpage 

\def\bibsection{}  
\centerline{ \textbf{\large  \textcolor{black}{Supplementary References } }}
\bigbreak

\bibliography{yanibib}